\documentstyle[aaspp4,psfig]{article}

\newcommand{\eg}{e.g.,~}

\newcommand{\etal}{et al.~}
\newcommand{\um}{\mbox{$\,\mu\rm m\,$}}
\def\mic{$\rm\mu m$}

\def\arcsec{''\hskip-3pt }
\def\arcmin{'\hskip-3pt }
\def\deg{\ifmmode^\circ\else$^\circ$\fi}

\def\lsun{{$\,{\cal L}_\odot$}}
\def\msun{{$\,{\cal M}_\odot$}}

\def\lfir{{$\rm {\cal L}_{\rm FIR}$}}

\def\R34{{$\rm f_\nu(60\um)/f_\nu(100\um)$}}
\def\R12{{$\rm f_\nu(12\um)/f_\nu(25\um)$}}

\def\wig#1{\mathrel{\hbox{\hbox to 0pt{%
  \lower.5ex\hbox{$\sim$}\hss}\raise.4ex\hbox{$#1$}}}}
\def\cre{$\rm CR\ e^-$}

\begin{document}
\title{Normal Galaxies in the Infrared\footnote{Lecture notes from the Summer School ``Infrared Astronomy: Today and Tomorrow,'' held in Les Houches, August 1998.  Editors F. Casoli and J. Lequeux.}\\}

\author{George Helou}
\affil{Downs Lab, 320-47,\\
       California Institute of Technology,\\
       Pasadena, CA 91125,\\
       USA}

%\begin{abstract}
%  Abstract goes here if needed
%\end{abstract}

\section{INTRODUCTION} 

A normal galaxy is one that derives its luminosity primarily
from nuclear burning in stars, and is at neither the high nor the low
extreme of the luminosity distribution.
In such galaxies, 
the visible and ultraviolet is the only direct window onto the 
photospheres
of stars, whereas the rest of the spectrum reflects reprocessed light from
stars (Figure 1).  Dust in the interstellar medium reprocesses part
of the stellar luminosity into infrared emission from a few \um to 1~mm or 
longer wavelengths.  The radio emission at millimeter wavelengths
derives from thermal emission from ionized plasmas in HII regions, 
whereas the synchrotron emission from cosmic ray electrons ($\rm CR\ e^-$) 
trapped in the magnetic field of the galaxy fill in the cm-wavelength
spectrum.  In the X-rays, the luminosity is dominated by very hot
plasmas at $\rm T\wig>10^6~K$ created by shocks from supernova
explosions.  The global spectrum of normal galaxies has some 
very stable signatures, and other quite variable aspects.
Much of this article is an attempt to understand what makes for
stability in the spectrum, and what significance to attach to
the variable parameters.

Since stars form in the interstellar medium (ISM), it is no surprise
that galaxies forming stars actively have substantial ISM luminosities,
conveyed overwhelmingly in the infrared, and often exceeding by 
large factors the optical and ultraviolet luminosity.
Studying galaxies in the infrared is thus tantamount to studying
the ISM and its properties, and star formation (SF) activity on large scales.

Astration, the cycle of star formation, nuclear burning and ISM
enrichment in heavy elements, is the primary secular change and 
main evolutionary process affecting the
chemical make-up and energy balance of the Universe.   Because the luminosity
function ({\it e.g.} Kim \& Sanders 1998) falls fast enough that the
integral over the population is dominated by the low end of the 
luminosity interval, over 90\%
of star formation in the local Universe takes place in normal disk galaxies, 
rather than the spectacular extremes.  Thus, normal galaxies
essentially cause and host Cosmic nuclear evolution.

Normal galaxies form a unique bridge between our detailed understanding of
the Milky Way and our global understanding of the Universe and its history.
Empirically, the integrated emission from galaxies must be compatible with
the properties of Galactic objects.  Our physical models of the local
energy balance and of large-scale processes in the ISM must be consistent
with the observed global behavior.  On the other hand, galaxy luminosity
and spectral properties are a crucial ingredient to modeling the observed
faint extragalactic source counts and infrared background.  Moreover,
modeling the counts and background requires more than a census of the
passive population ({\it cf.} Puget's article in this volume), and our
models of galaxy behavior must be consistent with population evolution
required to explain deep counts.  The enrichment in individual galaxies
must also add up to the observed current-day abundances.  Thus achieving an
accurate representation of normal galaxies would verify directly the
validity of underlying models of the Milky Way ISM, and is a necessary step
to the interpretation of cosmological results.

The InfraRed Astronomy Satellite (IRAS; Beichman \etal 1986) ushered in the
era of space-based infrared astronomy in a dramatic fashion, revealing a
stunningly rich infrared sky, unanticipated from the bits of infrared data
previously and heroically collected from the ground.  IRAS conducted the
first unbiased all-sky survey at 12, 25, 60 and 100\um, as well as a
spectroscopic survey with a resolving power of 20 between 7.5 and 23\um,
for objects brighter than 10 Jy or so.  IRAS had a profound influence on
astronomy in general, not just the infrared, because it represented such a
very large gain in sensitivity and spatial coverage, comparable perhaps to
going from attempting visual astronomy in daylight to observing in a dark
night (Beichman 1987; Soifer, Houcj \& Neugebauer 1987).  Another
significant factor in this influence was the  dissemination
of data products from IRAS, including source catalogs, image atlases and sky
brightness estimates, generated with great care and well characterized and
documented as to reliability, completeness, flux accuracy, and other statistical
attributes.

The Infrared Space Observatory (ISO, Kessler \etal 1996) took infrared
astronomy in space to new levels of sensitivity and into a new realm of
spectroscopy across the whole range of 3 to 200\mic.  Its versatile
instrumentation took to the rich and fascinating extragalactic infrared sky
first revealed by IRAS, and moved our knowledge of that sky an order of
magnitude further and deeper.

In what follows, methods traditionally used in the study of 
normal galaxies are described, and outstanding questions currently 
pursued in the field are stated.  The most prominent results
from the IRAS survey are reviewed in \S3.   Contributions
by ISO in the field of broad-band photometry are then presented (\S4),
followed by results in spectrospcopy (\S5).  Normal galaxy studies not
directly concerned with the ISM are reviewed in \S6.  The outlook and 
challenges in pursuing the interpretation of infrared data on the ISM are
discussed in \S7.

\section{THE STUDY OF NORMAL GALAXIES}

Normal galaxies span wide ranges of luminosity and many other parameters
such as density, intensity, metallicity, extinction, or light-to-mass
ratios.  This great diversity in properties, and the complex mix of
physical conditions within each system, have frustrated attempts at
deriving simple models for the behavior of these ``unremarkable'' galaxies.
Surveys of galaxies in the local Universe have thus become particularly
valuable for codifying the empirical evidence.  Survey data are typically
analyzed to generate statistical information or to address specific
questions, usually using global or integrated properties.  For instance,
luminosity functions are derived, or correlations between parameters are
established.  In such statistical studies, close attention must be paid to
the sample used, for its completeness, selection biases, parameter-space
coverage and other aspects could affect the results substantially.
Statistical results are usually interpreted in phenomenological terms,
explaining the observed trends for instance as reflecting the mixing of two
components with distinct properties.

A complementary approach is to study in great detail a few nearby, 
spatially well resolved systems, and to derive insight from the similarities 
or contrasts among these cases.  Such detailed case studies tend to 
search for physical insight by relating observational trends to the variation
of a physical parameter of the system.   While similar insights can also be 
pursued with statistical studies, case studies can make more convincing
arguments based on detailed physical models of the ISM on small scales
where homogeneity can be assumed.

Ultimately, one would like
all these diverse interpretations to be collected into a self-consistent
and complete picture.  The main obstacle to such unification is that each
observable $\rm P_{obs}$ is a complicated integral over the physical system:

$$\rm P_{obs}\ =\ \int _\Omega \int_{\sc l-o-s} f_P (U,n,T,X_P) d{\sc l} d\Omega,$$

\noindent where the integrals are taken over the solid angle of the galaxy and
through the line-of-sight respectively, and the contribution at each
point is a function of radiation intensity U, density n, temperature T,
and weighting functions $\rm X_P$ such as geometry or optical depth
effects; the form and dependencies of $\rm f_P$ 
vary greatly among observables.  While the availability of spatially resolved data 
simplifies the problem by narrowing down the solid angle and
therefore the variability of $\rm f_P$, it does not eliminate it entirely.

The empirical pursuits concerning normal galaxies typically aim at 
deriving a better description of the infrared properties, especially in terms
of identifying the ``fundamental parameters'' that stand behind the many 
correlated parameters, and pinning down the precise significance of observables
in terms of physical parameters.  Another main goal is to pin down
similarly the fundamental sequence defined by the evident progression
of infrared properties, and to understand its key significance.  Is the true
sequence simply defined by the infrared spectral energy distribution?
Is it driven by intensity variations?  Or by the UV content of the heating spectrum, 
and therefore the galaxy's content of young stars?  Are there secondary 
parameters defining truly significant multi-dimensional families of properties?
Can low-luminosity analogs be defined for objects with extreme properties,
making more cases available for study in the Local Universe?  An improved 
empirical understanding of the data in these terms will help improve our
understanding of extreme systems, help us better plan high redshift surveys,
and understand their biases, and will feed directly into deciphering the data on
the infrared cosmic background and on source counts at faint levels.

The physical understanding of star formation on the scale of galaxies must
start by addressing the episodic behavior of star formation, and the apparently
chaotic behavior on the kpc scale within disks.  Can
a cycle be identified with well defined phases?  Can those phases
be distinguished by observations, and can physical drivers and inhibitors
for episodes  be identified?  Can a galaxy then be understood in 
terms of a superposition of phases of many overlapping episodes, and can 
the factors be identified which regulate the ``steady-state'' of the whole disk
and the correlations between global parameters?  Ultimately, physical models
of star formation in galaxies would be used to simulate the history of
star fromation and chemical evolution in the Universe, and construct more
reliable models of primordial galaxies to guide the search for those 
first-generation objects.

\section{GALAXIES IN THE INFRARED:  THE IRAS ERA}

IRAS conducted the first unbiased all-sky survey
(or almost, covering 98\% of the sky), detecting
point sources down to 0.2~mJy or fainter at 12, 25 and 60\um, and down to 
1~Jy at 100\um (Beichman \etal 1986; Moshir \etal 1992).  IRAS also
carried out a spectroscopic survey with a resolving power of 20 between
7.5 and 23\um, for objects brighter than 10 Jy or so.  IRAS doubled the
amount of existing infrared data at $\lambda>5\um$\  within its first 
hour of observation.
The IRAS mission lasted ten months.  It resulted in a data set which
completely defined our knowledge of normal galaxies at $\lambda>5\um$\  until 
the launch of ISO.  The Kuiper Airborne Observatory (KAO) contributed its
share to the field, but because it was limited to high surface brightness
objects, its impact on normal galaxies remained minor.

IRAS measured fluxes for nearly 60,000 galaxies, perhaps half of which were
previously uncatalogued.  It allowed detailed studies of many nearby
galaxies, and of course of the Milky Way, establishing several crucial
connections between global properties of galaxies and specific aspects and
phases of the local ISM.  Some of the contributions of IRAS to normal
galaxies statistical properties are summarized in the following three
subsections.  See also the early review by Soifer, Houck \& Neugebauer
(1987), and the overviews in the 1991 Les Houches Summer School
proceedings. 

\subsection {Basic Parameters and Statitics}

\subsubsection {Infrared Luminosity}

A few key parameters have gained much currency in the study of the
integrated emission from galaxies in the infrared.  Foremost among these is
infrared luminosity $\cal L{\rm(IR)}$, used quite often as an indicator of
the total level of activity in the ISM.  Various authors have used
different spectral definitions for $\cal L{\rm(IR)}$, most commonly
favoring either the far-infrared (FIR) or the total infrared $\cal L{\rm
(TIR)}$.  The IRAS data naturally lead to the ``FIR'' definition (Helou
\etal 1988) of a synthetic band combining in a simple way the 60 and
100\um~ flux measurements.  \lfir\  in units of $\rm W~m^{-2}$ is defined by
$\rm {\cal L}_{FIR}~=~1.26~10^{-14}[2.58~f_\nu(60\mu m)+f_\nu(100\mu m)]$,
where $f_\nu(60\mu m)$ and $f_\nu(100\mu m)$ are in Jy.  Helou \etal (1988)
demonstrated that this combination approximates to within 1\% the flux in a
synthetic band with uniform transmission between 42.5 and 122.5\um for
blackbody and modified blackbody (with emissivity $\propto \lambda^n$)
curves with temperatures between 20 and 80~K, and for emissivity index $n$
between 0 and 2.  They argued this property should therefore also apply to
realistic spectral energy distributions of galaxies because those must be
made up of a superposition of modified blackbodies in this temperature
range.  It is essentially a coincidental result of the properties of the
IRAS filter shapes that the simple linear combination
allows us to estimate the luminosity in a well defined spectral window.
The real interest of FIR however is that this window encompasses a large
fraction, and therefore a representative measure, of the total infrared
luminosity.  Fortunately, as discussed in \S4 below, the ratio $\cal
L{\rm (TIR)/}L{\rm (FIR)}$ is on the order of 2, and varies relatively
slowly with the properties of galaxies.

While $\cal L{\rm (IR)}$ is a good indicator of the total luminosity from
the ISM of a galaxy, some authors have claimed it to be proportional to the
star formation rate.  This interpretation has been controversial, and is
most probably erroneous, as discussed in \S 3.4 below.  $\cal L{\rm (IR)}$
is an extensive quantity, best approximated as an integral over the
galaxy of the intensity of the interstellar radiation field
times the effective local optical depth of the dust to this radiation.  Its
interpretation as a measure of star formations is valid only when the
optical depth is high everywhere, and the radiation field is derived
primarily from young stars.

The normal galaxies under discussion have $\cal L{\rm (IR)}$ in the range
$10^7-10^{11}$\lsun.  Over this interval, the distribution of infrared
luminosities is well described by a power-law function with an index in the
range -2 to -2.5 (Kim \& Sanders 1999).  In ultra-luminous galaxies 
(Houck \etal\ 1985, Harwit \& Houck 1987; Sanders \& Mirabel 1996), 
$\cal L{\rm (IR)}$ exceeds
$10^{12}$\lsun, emanating from an ISM heated by prodigious star formation
or possibly by an active galactic nucleus.  At the other extreme of {\cal
L}(IR), IRAS showed that a minimum
luminosity in the mid-infrared can be expected from the photospheres of
stars in galaxies (Soifer \etal 1986), but was not sufficiently sensitive
to establish whether galaxies emit a minimum luminosity in the far infrared
in addition to the photospheric emission, and apart from the ISM emission
found even in Elliptical galaxies (Jura 1986; Knapp \etal 1989).  Even ISO
data may not be adequate to address this question.

\subsubsection {The Infrared-to-Blue Ratio}

The infrared-to-blue ratio (IR/B) compares the luminosity reprocessed by
dust to that of escaping starlight, and is therefore short-hand for the
ratio of total infrared to total emerging photospheric emission, from the
near-infrared to the ultra-violet.  This ratio ranges broadly, from $<0.01$
to $\sim100$ in known galaxies (Figure 1), with a weak dependence on
parameters such as morphology (Sauvage \& Thuan 1994), suggesting that it
fluctuates substantially in time for the same galaxy.

Because it is a dust-to-star luminosity ratio, IR/B might be interpreted 
as a function of the effective optical depth of a simple system of stars and dust,
with $\rm IR/B\propto[\tau-(1-e^{-\tau})]/(1-e^{-\tau})$.  Most normal
galaxies however are far from this simplicity:\   Some of the visible luminosity
comes from stars not participating in the heating of dust, especially old stars
in the bulge or far from interstellar clouds. On the other hand, some of the
infrared luminosity is traceable to stars completely hidden inside dense clouds,
and all but invisible from the outside.  Still,
IR/B could be turned into a measure of effective optical depth using assumptions
for the geometry of stars and dust, for the spectral energy distributions of heating
stars, and for the dust efficiency of absorption as a function of $\lambda$ 
(e.g. Xu \& Helou 1996).  Such assumptions may be well bounded in special cases
such as specific parts of a disk, or star-burst galaxies, but their uncertainty 
is harder to estimate for whole galaxies.

While the interpretation above may be a good approximation for IR/B$\wig>$1,
an alternative at lower values is that IR/B  characterizes the ratio of 
current or recent star formation rate to the long-term average rate.
In a picture where geometry, heating spectra and dust properties are fixed,
the dust luminosity may be interpreted as a measure of stars interacting
with the ISM, and thus of the stars recently formed out of that ISM.
Unfortunately, the precise time intervals implied by the terms ``recent'' 
and ``long-term'' are themselves a function of the history of star formation 
in a given galaxy.  This interpretation of IR/B applies to systems whose
infrared emission is limited by the amount of heating photons available,
whereas the optical depth interpretation applies to systems whose infrared
emission is limited by the amount of dust available to heat.  A dwarf galaxy
like NGC~1569 for instance may be undergoing intense star heating by young
stars, but have very little neutral ISM left, resulting in low IR/B.
On the other hand, a quiescent galaxy may generate most of its infrared emission
in HI clouds heated by the older stellar population, and display a similarly
low IR/B.  The degeneracy between these two cases can only be broken by use
of other observables, such as stellar population and morphology, infrared
spectral energy distribution, or spectroscopic diagnostics of the gas.

The large dispersion in the IR/B ratio points to dramatic fluctuations in the
star formation activity during the lifetime of a galaxy, hence the notion that
much of star formation is episodic.  Small irregular galaxies are 
characterized almost entirely by intense episodes of star formation separated
by long quiescent periods.  There is abundant evidence that galaxies undergoing
a nuclear starburst cannot sustain it for more than a small fraction of
their lifetime (Sanders \& Mirabel 1987).  
Large disk galaxies in steady state appear to have several episodes
under way at any time within their disk, and observed parameters integrated over
the whole galaxy are an ensemble average across all phases of these episodes.

\subsubsection {IRAS Colors}

Mid- and far-infrared color ratios describe the shape of the dust emission at 
those wavelengths.  To first order, the spectra of normal galaxies are organized 
into a single family of curves, manifested as a linear locus that galaxies
occupy in the IRAS color-color diagram (Helou 1986).  This diagram was quite
surprising at first look, since it implies that the spectrum becomes cooler
when judged by the mid-infrared  R(12,25)=$\rm f_\nu(12\um)/f_\nu(25\um)$
color as it gets warmer in the far-infrared
color ratio R(60,100)=$\rm f_\nu(60\um)/f_\nu(100\um)$ (Figure 2).  
It turns out that this behavior results from the interplay
of two spectral components: blackbody-like emission from classical grains in
temperature equilibrium mostly at the longer wavelengths, and relatively
fixed-shape mid-infrared emission from tiny grains with a few hundred atoms or less, 
intermittently heated by single photon events (more in \S 4.2).

The seqence of infrared colors in Figure 2
was clearly associated with a progression towards greater dust-heating intensity,
as illustrated by the progression of colors in the California Nebula as one
approaches the heating star (Boulanger \etal 1988).  The cool end of the color
sequence corresponds to cool diffuse HI medium and to quiescent molecular clouds,
whereas the warm end corresponds to the colors of HII regions, star-bursts and
galaxies with high IR/B ratios and higher infrared luminosity.  
It is thus natural
to associate the color progression with a sequence of star formation activity 
in galaxies, signalling in simplest terms an increase of the fraction of
$\cal L{\rm(IR)}$ traceable to young stars (Helou 1986; \S 3.4) 

The span of this activity sequence corresponds to a significant shift in the
infrared specturm.  In the coolest spectra, the peak in $\rm f_\nu$ is at
$\sim 150\um$; it shifts to $\sim 60\um$ or shorter in the warmest galaxies.  
However,
in spite of their usefulness as indicators of ISM activity, the color ratios 
derived from IRAS data are useless for deriving dust temperature, which would
then be used to derive a dust mass.  The infrared spectra of galaxies
are a weighted mean over
a broad range of environments in which dust emission arises, and are driven 
far from blackbody temperatures by non-equilibrium emission from tiny grains
which fluctuate between near zero K and high excitation by single photons
(\S 3.4 and \S 4.2).

\subsubsection {Other Estimators}

Of many other parameters used in analyzing the empirical properties
of galaxies from IRAS data, only two are shown here, for illustration rather
than completeness.  In using such estimators, one should be aware of the
assumptions that enter into relating them to specific physical quantities,
and estimate the uncertainties resulting from the inaccuracy of the
assumptions in addition to the inherent measurement uncertainty.

The ratio of $\rm f_\nu(100\um)$ to 
the HI $\rm \lambda~21cm$ line flux has been used as
an indicator of the dust-to-gas ratio.  This interpretation however works
only in very broad terms when comparing galaxies, since several factors
other than dust to gas ratio affect the parameter (Helou 1985).  
For instance, the 100\um-to-HI ratio
scales at least linearly with heating intensity, and depends on the assumed 
ratio of atomic to molecular hydrogen, as well as on the fraction of HI-related
dust which is heated sufficiently to emit at 100$\um$, as opposed to dust in
an extended HI envelope which remains cold enough as to be irrelevant at 100$\um$.
This estimator works better locally within galaxy disks as a ratio of
surface brightness at 100\um\  and in HI as opposed to a global parameter 
(Dale \etal 1999).  See Melisse \& Israel (1994) for an interesting analysis.

The ratio of FIR flux to CO flux has been used as a measure of star formation
efficiency, in the context of interpreting $\cal L{\rm (FIR)}$ as a measure of 
star formation rate, and $\cal L{\rm (CO)}$ as a measure of the gas mass 
available for star formation.  The 
inverse quantity can also be interpreted as an indicator of how long the
current star formation rate can be sustained by the system.  The caution here is
that this estimator is useful only to the extent allowed by the uncertainties 
in interpreting $\cal L{\rm (FIR)}$ and $\cal L{\rm (CO)}$ as just stated.
$\cal L{\rm (CO)}$ in particular traces CO
rather than total $\rm H_2$ mass, and includes a significant dependence 
on the CO excitation conditions,
as well as optical depth effects (Maloney \& Black 1988, Aalto \etal\  1994).

It is clear that much caution needs to be exercised in using all these
estimators.  Using several simultaneously helps in narrowing down the
uncertainties of interpretation.  For example, Knapp \etal (1987) combined
$\rm f_\nu(100 \um)$/F(HI) with $\rm f_\nu(100 \um)$/F(CO) to trace the
origin of the infrared emission.  The discussion above illustrates the
importance to the study of normal galaxies of understanding ISM physics and
star formation processes on scales from a parsec to a kpc.  This
understanding must be derived from empirical and physical modeling studies
of the Milky Way and Local Group galaxies, at the best available spatial
and spectral resolution.

\subsection {Correlations}

Correlative analysis is a commonly used tool in the
study of normal galaxies, and provides a more powerful tool than analysis of
individual parameters, as illustrated in \S 3.1.4.  
Correlations however may be easier to establish than to interpret correctly.  Many
correlations are well known between various indicators of star formation
activity, for instance IR/B, R(60,100) and $\cal L{\rm (FIR)}$ are all
positively correlated, to an extent that varies with the sample used (Soifer \&
Neugebauer 1991; Bothun \etal 1989).  In
principle these correlations contain information on geometry and physical
parameters such as density and heating intensity, and on the manner in
which star formation affects these parameters, is affected by them, and
whether there are balancing forces.  In practice however, it can be very
difficult to derive such inferences.

Particularly easy to generate are correlations between extensive parameters,
those which scale with the extent of a system.  Luminosity is such an extensive
parameter, scaling with the square of
the distance to the galaxy, as opposed to color ratio for instance, which is
distance independent.  Correlations between
extensive parameters are of limited interest because they tend to be exaggerated
by the distance scaling, since errors on the distance will affect all such
estimators equally, reinforcing the appearance of a positive correlation.  
Such correlations are also exaggerated by the spread
of system sizes, suggesting great significance, whereas the main information
content is that all extensive quantities tend to be greater in larger systems.
The accepted procedure to avoid such vacuous correlations is to normalize
extensive quantities by a system size parameter such as luminosity or mass,
thereby reducing as many parameters as possible to distance-independent 
expressions.  Examples of correlations and their interpretation appear in \S 3.3
and in \S 5.

\subsection {The Infrared-Radio Connection}

The tightest and most universal correlation known among galaxy fluxes
connects the FIR emission from dust heated by stars with 
the non-thermal radio emission in the range from 2 to 50 cm, 
which is known to be synchrotron
radiation from \cre\  trapped in the interstellar magnetic field.
This tightest of correlations is also the most puzzling.  First, because of the 
indirect connection between the two mechanisms and populations of emitters,
and the many parameters involved in producing the luminosities at the two
wavelengths, such as current stellar population properties, dust optical depth, 
magnetic field strength, and \cre acceleration and escape mechanisms.
Secondly, surprising because of the large difference between the two luminosities;
the ratio between the two bands is about $5\times 10^5$, with a dispersion of about
50\%.  See reviews by Helou (1991) and Condon (1992).

The first mention of a close relation between 
far infrared and radio emission appeared as soon as the relevant data 
became available.  Rickard and Harvey (1984) pointed out a strong 
correlation in a sample of 30 late type galaxies between the central 
emission at 20 cm and the emission in the 40 to 160~${\mu}$m spectral range.
They related the correlation to star formation activity, and assumed
the non-thermal radio emission to be dominated by supernova remnants,
as predicted by Harwit and Pacini (1975).
Rickard and Harvey (1984) were puzzled however by indications 
that the same correlation applied to the disk emission, because that 
would imply cooperation between the magnetic field $B$, the interstellar 
gas density $n$, and the density of \cre.
Since 1984, the vastly superior data returned by the VLA and IRAS 
have placed the correlation on a far more solid empirical footing
without changing the basic facts. The situation on the side 
of interpretation however has improved more slowly, 
and a consensus has yet to emerge as to the physical significance 
of this correlation.

The global correlation was noticed 
early on in the IRAS data by Dickey and Salpeter (1984), then independently 
established by de Jong \etal (1985) using a sample of IRAS sources 
selected from the early mission returns, and by Helou, Soifer and 
Rowan-Robinson (1985) using an optically selected list of galaxies 
in the Virgo cluster and in the field.  Regardless of the sample selection 
criteria, the ratio $Q$ of infrared to radio in
samples of star forming galaxies displays an intrinsic population
dispersion of 50\% or less (Helou,
Soifer and Rowan-Robinson 1985; Sanders and Mirabel 1985; Condon and
Broderick 1988; Wunderlich and Klein 1988; Unger {\it et al.} 1989)
over four decades in luminosity, though non-linearities have been claimed,
with the radio increasing faster than the infrared power (Cox \etal
1988; Menon 1991).  The correlation has been shown to hold at high redshifts
(e.g. Karoji, Dennefeld and Ukita (1985), Hacking \etal 1989)
and has become an accepted property of all star forming galaxies (Condon 1992).

A definitive empirical treatment of the correlation was published by 
Condon, Anderson \&  Helou (1991), showing that it is asymptotically linear
as the ratio of radio-to-visible, or equivalently of infrared-to-visible,
increases, so that the ``true'' correlation is evident when the system
is powered by young stars (Figure~3; see also Xu 1990).  This happens consistently
for galaxies with infrared-to-visible luminosity ratios greater than $\sim0.3$, 
which are still relatively optically thin galaxies.  For galaxies less
active in star formation, $Q$ rises slowly above the standard correlation value.
Within galaxy disks,
the infrared emission is clearly more centrally peaked than the radio 
emission (Marsh \& Helou 1998; Marsh \& Helou 1995; Bicay \& Helou 1990;
Rice \etal 1990; Beck \& Golla 1988; Wainscoat \etal 1987).

The theoretical understanding of this correlation is still imperfect.
The physical explanation needs to invoke a two-part argument, namely
a luminosity balance to explain optically thick systems, and a filtering 
match to explain optically thin systems.  In an optically
thick galaxy, all dust-heating radiation is re-radiated in the infrared
on the one hand, and all \cre\  created are trapped by the magnetic fields 
long enough for their available energy to dissipate as synchrotron 
luminosity.  In this case, linking the dust-heating
luminosity and the \cre\  luminosity is sufficient to produce the observed 
correlation.  This linkage is achieved because stars more massive than
$\sim 8$\msun dominate the dust heating {\it and} produce Type~II supernovae
whose shocks accelerate cosmic rays including \cre\  (V\"olk 1989).
Condon (1992) has shown that the ratio of heating to \cre\  luminosities is 
not very sensitive to upper mass cut-off of the Initial Mass Function (IMF), 
nor to age of starburst.  See for historical interest also Lequeux (1971),
Klein (1982) and Kennicutt (1983).

For optically thin galaxies, the effective dust optical depth of the galaxy
to heating radiation must match its efficiency at extracting synchrotron
radiation from \cre.  Helou \& Bicay (1993) presented a model which achieves 
that match by assuming simple connections among physical parameters in 
the ISM, most importantly between the density of the medium and the magnetic
field intensity.  The model takes into consideration geometry of dust
and stars and magnetic field, diffusion, radiative decay and escape of \cre,
and develops a picture where the radio disk is a smeared version of the
infrared disk, the smearing being greater for more transparent disks.
This model was found subsequently to be consistent with the observed details
of the infrared-radio correlation within disks of galaxies (Marsh \& Helou 
1998).  However, models for the physics underlying the correlation remain 
sufficiently complex that other mechanisms
for the radio emission have been proposed because they provide closer
connection to the infrared (\eg Harwit in this volume).

\subsection {The ``Two-Component Model''}

A ``two-component model'' is any representation of the spread of properties
of a system as the result of the superposition of two components
whose properties define the extremes of that spread.
Such a model was invoked to explain the IRAS color-color diagram,
describing normal galaxies as a linear combination of a quiescent
component and a star-forming component, whose mixing ratio determines the
colors of a given system (Helou 1986).  
The FIR-cold, ``cirrus-like'' component
is supposed to arise primarily from low density, low radiation intensity
quiescent regions heated primarily though not exclusively by older stars.
The FIR-warm ``active'' component corresponds to dust heated in the vicinity
of star-forming regions.  Each of these components has its own luminosity 
and effective optical depth, and one could in principle solve for at least
some of these quantities in any given galaxy with sufficient data.  
The infrared luminosity and optical depth of the active component combine to yield
the heating luminosity in star forming regions, and therefore the star formation
rate.  This type of decomposition is mostly morphological, and particularly
useful for nearby, well resolved galaxies such as M31 (Xu \& Helou 1996).
Note the similar but lower dynamic range decomposition proposed by Larson \&
Tinsley (1978) using the visible colors U-B and B-V.

A more physical and useful decomposition would be to represent the infrared
luminosity as $\rm {\cal L}_{ION}(IR)~+~{\cal L}_{NON}(IR)$, where the
first term reflects heating by ionizing stars, and the second heating by
non-ionizing stars.  The very existence of $\rm {\cal L}_{NON}(IR)$ has
been challenged by Devereux \& Young (1990, 1992, and subsequent papers).  
Their arguments however are
critically dependent on very uncertain assumptions about the upper mass
cut-off of the stellar mass function in galaxies, and can therefore be
easily dismissed.  Furthermore, detailed studies of nearby galaxies
prove that $\rm {\cal L}_{NON}(IR)$ can contribute more than half $\cal
L{\rm (IR)}$ (Walterbos \etal 1987; Rice \etal 1990; Rand \etal 1992; Xu \&
Helou 1996).  Smith \etal (1991) show in NGC~4736 a striking example of the
infrared emission from the nuclear region being dominated by $\rm
{\cal L}_{NON}(IR)$, and surrounded by a star forming ring whose emission is
dominated by $\rm {\cal L}_{ION}(IR)$.

How can the mixing ratio of the two components in this physical
decomposition be estimated?  Given its definition, the best indicator would
be a measure of the hydrogen recombination rate and thus of the total
ionizing flux in the system, preferably obtained from a long-wavelength
transition such as Br$\gamma$ to avoid extinction effects.  That would
determine the amount of ionizing flux and the total stellar luminosity, 
from which the dust heating could be
estimated, yielding $\rm {\cal L}_{ION}(IR)$.  Another approach to
estimating the mixing ratio is offered by the infrared-radio correlation.
Because of the close coincidence between the lower mass limit for ionizing
stars $\rm\sim6$\msun, and the lower mass limit for supernova-capable
stars $\rm\sim8$\msun, one could associate $\rm {\cal L}_{ION}(IR)$ with
the radio-loud component and thus use $\rm {\it Q}={\cal L}(IR)/{\cal L}(radio)$
as an index to $\rm {\cal L}_{ION}(IR)$/$\rm {\cal L}_{NON}(IR)$.

In any case, galaxies at the extremes of the $\rm {\cal L}_{ION}(IR)$/$\rm
{\cal L}_{NON}(IR)$ can be readily identified since parameters such as
IR/B, {\cal L}(FIR)/{\cal L}(HI), IRAS colors or $Q$ also approach their extreme
values in such galaxies.  More detailed modeling as in papers mentioned earlier
might be needed to estimate the mixing ratio in specific galaxies.  
On the other hand, infrared luminosity and
morphology would be poor indicators of a galaxy's position on that mixing
ratio scale.

The two-component model proposes a simple picture where the infrared
properties of a galaxy are determined by the mixing ratio $\rm {\cal
L}_{ION}(IR)$/$\rm {\cal L}_{NON}(IR)$, and where each of these infrared
components results from a heating luminosity and a corresponding optical
depth.  While the mixing ratio in a given galaxy may be quite uncertain,
one can select samples of galaxies for statistical analysis where this ratio
is biased towards $\rm {\cal L}_{ION}(IR)$\  or $\rm {\cal L}_{NON}(IR)$.
Studies of the star formation rate in galaxies should use samples
biased towards high mixing ratios of ionizing to non-ionizing luminosities,
to support the simple assumption that {\cal L}(IR) is a good measure of the
star formation rate.  Infrared selected samples such as the IRAS Bright
Galaxy Sample (Soifer \etal 1989) are $\sim80\%$ populated by galaxies
dominated by $\rm {\cal L}_{ION}(IR)$, and comprise $\le10\%$ objects
dominated by $\rm {\cal L}_{NON}(IR)$.  Objects with high mixing ratios
also tend to have high IR/B ratios, since the optical depth associated with
star formation naturally tends to be quite elevated.  On the other hand,
studies aimed at the structure of stars and the ISM in galaxies should use
broader samples, such as optically selected or volume-limited samples.
Optically selected samples such as one derived from the Uppsala Galaxy
Catalogue (Bothun \etal 1989) have 40 to 50\% of galaxies dominated by $\rm
{\cal L}_{ION}(IR)$ and 20 to 25\% dominated by $\rm {\cal L}_{NON}(IR)$.
Volume-limited or cluster samples may be even richer in quiescent galaxies.
In galaxies with low mixing ratios, the optical depths are harder to
estimate, but remain smaller than unity in general.

\section {ISO RESHAPES THE DUST CONTINUUM}

In assessing the impact of
ISO on our knowledge of the continuum emission from galaxies, the most 
exciting developments to date have touched the mid-infrared, as 
might be expected
from ISO's substantial gains in sensitivity and spatial resolution
at those wavelengths.  Gains at
the longer wavelengths are more subtle.

First, the various
surveys conducted by ISO are reviewed, then new results in the mid-infrared
spectral range are described, in the areas of spectro-photometry and 
broad-band colors, and implications for the overall spectral
energy distribution are mentioned.  Finally, spatially resolved studies 
of galaxies at wavelengths available to ISO are presented.

\subsection {ISO Surveys of Galaxies}

While ISO was an observatory rather than a survey mission, many surveys
were carried out using various capabilities of its versatile payload.
Some prominent surveys that concern normal galaxies either directly
or indirectly are listed below.  This is by no means an exhaustive list,
especially since most ISO data have yet to be published.

\begin{enumerate}
\item Mid-infrared maps of nearby galaxies were obtained under the ISO-CAM
(C. C\'esarsky \etal 1996) guaranteed time (GT) program, targeting large
angular-size galaxies in various categories, such as early-type, spirals
barred and non-barred, dwarf irregulars, and active (Vigroux 1997).  There
were also surveys of galaxies in Virgo, Coma and other clusters.  
All galaxies
were surveyed in the LW2 (6.75$\um$) and LW3 (15$\um$) filters, and some in
other filters within the 3 to 18$\um$ wavelength range of ISO-CAM.  In
addition, several were observed with the Circular Variable Filter (CVF),
which yields images at a spectral resolution of about 20 over most of the
same wavelength range.  These data were taken mostly with $3\arcsec$~ pixels, with
an effective resolution of 7 to $9\arcsec$~ half-maximum width.

\item Far-infrared spectral surveys were
carried out under the GT program of the ISO-LWS (Clegg \etal 1996),
most notably for a sample of infrared-bright galaxies, meaning those
with a flux density greater than 50~Jy at 60$\um$, and of ultra-luminous
galaxies (Fischer \etal, 1999).  Most objects were
observed with a LWS low-resolution full spectral scan covering 45 to $195\um$.

\item Far-infrared maps of well-resolved nearby galaxies were obtained
under the ISO-PHOT (Lemke \etal 1996) GT program at 60, 100 and $175\um$,
most notably of M31, M33 and M101.

\item Photometry at $\lambda\ge60\um$ was also carried out under the
ISO-PHOT GT program for several samples, including 75 bright (B$<$12~mag)
galaxies from the Revised Shapley-Ames Catalog, and selected objects in the
Virgo Cluster.  These samples were observed (PI R. Joseph)
at 60, 100 and 175$\um$, as
well as 12$\um$ (ISO-CAM filter LW10), with additional data collected from
the ground in the near-infrared and the submm.  The purpose was to extend
the spectral energy distribution, look for cold dust, and investigate
nuclear activity.

\item The surveys above were coordinated by J. Lequeux, resulting in several 
coherent data sets of great interest to normal galaxy studies.
The imaging with CAM at 6.75 and 15\um\, sampled galaxies at different distances,
from the Magellanic Clouds out to the Virgo Cluster, under programs {\tt``CAMSPIR''}
and {\tt ``VIRGO''}; program {\tt ``CAMSFR''} imaged star-forming regions 
at additional wavelengths, and with the CAM-CVF mode.  In Virgo, at least
thirty galaxies were imaged with CAM, and measured with PHOT at $\lambda\ge60\um$,
while their $\lbrack$CII$\rbrack$ ~$\lambda 157.7\um$ line  
was targeted with LWS (PI K. Leech).

\item Open time projects included several galaxy surveys, such as the 
Knapp \etal (1996) 
study of early type galaxies, the Lu \etal study of infrared-cold galaxies,
and the Metcalfe \etal (1996) BCD/Irr survey. 

\item The ISO Key Project on the Interstellar Medium of Normal Galaxies
(Helou \etal 1996) under NASA GT collected data on a set of sixty galaxies
that explore the full range of morphology, luminosity, infrared-to-blue
ratio and far-infrared color among star-forming galaxies.  These sixty
objects were selected to be small in their IRAS emission size compared to
the $80\arcsec$~ LWS beam and the $3\arcmin$~ ISO-CAM field of view, so as to
allow studies of their global properties.  In addition, nine nearby
galaxies were mapped to the extent possible, including NGC~6946, NGC~1313,
IC~10, and parts of M~101.  For most galaxies, maps were
obtained at 7 and $15\um$ with ISO-CAM, spectro-photometry was obtained
with ISO-PHOT-S between 3 and $12\um$, and far-infrared
fine-structure lines were targeted with ISO-LWS, attempting to measure as
many as possible of the following lines, in the order listed:
$\lbrack$CII$\rbrack$ ~$\lambda 157.7\um$,  
$\lbrack$OI$\rbrack$  ~$\lambda 63.2 \um$, 
$\lbrack$NII$\rbrack$ ~$\lambda 121.9\um$,  
$\lbrack$OIII$\rbrack$~$\lambda 88.4 \um$, 
$\lbrack$NIII$\rbrack$~$\lambda 57.3 \um$,  
$\lbrack$OIII$\rbrack$~$\lambda 51.8 \um$,
$\lbrack$NII$\rbrack$~$\lambda 145 \um$.

\item The ISO-PHOT Serendipity Survey gathered data during  satellite
slews between target observations with the $170\um$ channel.  By the end of
the mission, data had been collected over $150,000\deg$ of slew track, with
an estimated 4,000 galaxies detected (Stickel \etal 1998).  This data set
will be a unique source of far-infrared fluxes for thousands of galaxies
with IRAS detections at $f_\nu(100\um)\wig>2$~Jy.

\item By its nature as an observatory-class mission, ISO has generated 
a rich archive containing all the observations of individual galaxies,
groups, or clusters of galaxies investigated by various observers for
specific questions.  This collection constitutes a {\it de-facto} survey of
unique or peculiar objects from which one could learn much about the less
exotic cases ({\it e.g.} Smith 1998; Smith \& Madden 1997; Lu \etal 1996;
Jarrett \etal 1999; Valentijn \etal 1996; Xu \etal 1999).  
Many useful survey samples
can also be constructed after the fact by selecting objects out of the ISO
archive once it becomes available in the summer of 1999.
\end{enumerate}

Though not addressed directly in this review, ground-based infrared surveys
will make fundamental contributions to our view of normal galaxies,
especially the Two-Micron All-Sky Survey (2MASS, Skrutskie \etal
1999) and Deep European Near-Infrared Survey (DENIS, Epchtein \etal 1999),
Apart from these large systematic surveys, 
several near-infrared imaging surveys of nearby galaxies are
already revealing some surprising results.  Grauer \& Rieke (1998) for
instance  demonstrate that spiral arms are almost as contrasted in the K
band as they are in the B band.  See also Terndrup \etal (1994).

\subsection{Mid-Infrared Spectra}

In the integrated spectra of galaxies, the mid-infrared marks the
transition from emission dominated by stellar photospheres to re-radiation
by interstellar dust.  ISO has shown the details of this transition for
the first time by providing continuous coverage, and filling in some
crucial details as discussed in this and the next two sections.  
The
relevant data were acquired with ISO-PHOT, PHT-S module from 2.5 to 5 and
from 5.7 to 11.6\um\ (Lemke \etal 1996); with ISO-CAM in the CVF (Circular
Variable Filter) mode from 5 to 16.5 (C. C\'esarsky \etal 1996); and with SWS
from 2.5 to 45\um\ (de Graauw \etal 1996).

The transition from stellar to interstellar emission is well illustrated by
the spectra of Virgo Cluster galaxies collected by Boselli \etal (1998).  Its
precise location and therefore the interpretation to attach to mid-infrared
fluxes can be parametrized by a ratio such as IR/B (\S 3.1.2).  
Interstellar dust emission takes over by
5\um when this ratio exceeds 0.5, and at shorter wavelengths for higher
ratios.  As might be expected, Elliptical galaxies are dominated by stellar
emission, both photospheric and from circumstellar dust shells, and
therefore provide the templates that one subtracts to isolate the
interstellar emission component in Spiral galaxies (Boselli, Lequeux \&
Contursi 1997; Madden, Vigroux \& Sauvage 1997).  In addition, ISO
sensitivity has allowed us to study the small amounts of 
ISM contained in Elliptical galaxies, and to look into differences
with the ISM of Spirals (Knapp \etal 1996; Fich \etal 1999, Madden \etal 1999,
Malhotra \etal 1999; note also the study of E+A galaxies in
the Coma Cluster by Quillen \etal 1999).

\subsubsection{The Aromatic Features}

IRAS data had already indicated (Beichman 1987; Puget \& L\'eger 1989;
Boulanger \& Cox in this volume) that
the mid-infrared emission from the ISM was dominated by small fluctutating
grains and Aromatic Features in Emission (AFE).  ISO has not only
established into fact what had been hypothesis, but is also
allowing us to address quantitatively the mid-infrared energy budget across
various emission components, and to investigate the variation of this
budget from galaxy to galaxy.  ISO data are generally consistent with
older data from ground observations, including early M82 spectra by 
Willner \etal (1977), ground-based
surveys (Roche \etal 1991), and IRAS-LRS data (Cohen \& Volk 1989).

The AFE appear in two main groups, one stretching from 5.5 to 9\um, with
peaks at 6.2, 7.7 and 8.6, and the other one starting at 11\um\ and
extending to 12.5\um\ (Figure~4; Lu \etal 1999; Helou \etal 2000).
The shape and relative strengths of the features are quite similar to
``Type A sources'' which are the most common non-stellar objects in the
Milky Way: reflection nebulae, planetary nebulae, molecular clouds, diffuse
atomic clouds, and HII region surroundings  (Geballe 1997, Tokunaga 1997, 
and references therein).  Quantitatively
similar spectra have been reported from spectroscopic observations with 
PHT-S, ISO-CAM CVF or ISO-SWS 
on a variety of Galactic sources (Roelfsema \etal 1996;
Verstraete \etal 1996; C\'esarsky \etal 1996a; 1996b; Boulanger \etal 1996;
Mattila \etal 1996; Beintema \etal 1996; Uchida, Sellgren \& Werner 1998)  
and a number of galaxies (Boulade
\etal 1996; Vigroux \etal 1996; Acosta-Pulido \etal 1996; Metcalfe \etal
1996).  ISO-SWS spectra
with greater spectral resolution show AFEs with the same shape, 
a clear indication that they are spectrally resolved by PHT-S
data at a resolution of $\sim20$.

There is good evidence linking the AFE to Polycyclic Aromatic
Hydrocarbons (PAH), but no rigorous spectral identification of specific
molecules (Tielens 1999; Puget \& L\'eger 1989; Allamandola
\etal 1989).  It is generally agreed that the emitters are small
structures, no more than a few hundred atoms, transiently excited to high
energy levels by single photons.  The relative fluxes in individual AFE, 
and the general shape of the specturm, depend very weakly on
galaxy parameters such as the far-infrared colors (Figure~5).
This is direct evidence that the emitting particles are not in
thermal equilibrium.  As an estimate of the AFE relative strength, 
the integrals from 5.8 to 6.6\um, 7.2 to 8.2\um, and 8.2 to 9.3\um \ 
are in the ratio 1:2:1 in the PHT-S spectra obtained under the 
Key Project on Normal Galaxies (Helou \etal 2000).
The strongest detectable variation in the same data set is a 
slightly stronger 11.3\um AFE in the colder galaxies (Lu \etal 1999).
Since this
feature is linked primarily to neutral PAHs as opposed to the shorter
wavelength features which have a strong ionized PAH contribution (Tielens,
1999), one is tempted to interpret this trend as a result of the
more active galaxies having a greater contribution to their luminosity
originating in regions heated by harder radiation fields.  

An important consequence of the invariant shape of the spectrum up
to 11\um\,, even as the infrared-to-blue ratio reaches high values, 
is that the 10\um\, trough is best  interpreted as a gap between
AFE rather than a silicate absorption feature.  An absorption feature would
become more pronounced in galaxies with larger infrared-to-blue ratios, and
that is not observed (Sturm \etal 2000).

The fraction of starlight processed through AFE has been under debate since
the IRAS mission (Helou, Ryter \& Soifer 1991), and can now be directly
estimated using the new ISO data for the sample described above.  
AFE account for about 65\% of the total power within the 3 to 13\um\ range, and
about 90\% of the total power in the 6 to 13\um\ range.
% and for about 45\% of the flux within the 12\um\ IRAS band.   
The AFE between 6 and 13\um\ carry  25 to 30\% of {\cal L}(FIR) in 
quiescent galaxies, or  12\% of the total infrared dust
luminosity between 3\um \ and 1~mm, whereas all ISM emission  at
$\lambda<13\um$ comes up to $\sim18$\% of the total dust emission.
The ratio AFE-to-FIR gradually drops to less than 10\% 
in the most actively star forming
galaxies, i.e. those with the greatest {\cal L}(IR)/{\cal L}(B) ratio or IRAS color
$R(60/100)$, following the trend already noted in Helou, Ryter \& Soifer
(1991).  
Boselli \etal (1997 and 1998) have interpreted
similar trends as evidence for the destruction of AFE carriers in more
intense radiation fields.  The 3.3\um\, feature carries about 0.5\%
of the total AFE luminosity longwards of 5\um\,, a significantly smaller
value than that reported by Willner \etal (1982) for M~82.

\subsubsection {The Mid-Infrared Continuum}

In addition to the AFE, there is clear emission bridging between them,
even in the 10\um\, trough.  This emission may well be related to the
AFE carriers, since its shape is constant, scaling with the AFE strength.
Boulanger \etal\  (1998) have discussed this emission in terms of
Lorentzian wings to the AFE.

Less energetic but more surprising is the  continuum
detected in ISO-PHT-S data (Helou \etal 2000)
shortward of 5\um\, (see for instance the model spectra of D\'esert \etal 1990).
This unexpectedly strong 4\um\, 
continuum flux density is positively correlated with the AFE
flux, strong evidence linking it to dust rather than stellar
photospheres.  It appears to follow a power law $f_{\nu}\propto\nu^{+0.65}$
between 3 and 5\um, with an uncertainty of 0.15 on the power-law index.
However, the flux density $f_{\nu}$ around 10\um\ is three times higher
than the continuum level extrapolated to 10\um\ from the spectral shape
between 3 and 5\um, leaving open the nature of the connection between the
4\um\, continuum and the carriers of the AFE.  Bernard \etal (1994) have
reported evidence for continuum emission from the Milky Way ISM in
COBE-DIRBE broad-band data at these wavelengths, with comparable amplitude;
ISO however provides the first clear detection of the spectral shape of
the emission.

Extrapolating the continuum from the 
3 to 5\um\, range out to longer wavelengths, and assuming the AFE are 
superposed on it, one finds that the
continuum contributes about a third of  the luminosity between 3 and 13\um,
and 10\% between 6 and 13\um, the
balance being due to AFE and associated bridge emission.  
Against this extrapolated continuum, the
AFE, defined again as the emission from 5.8 to 6.6\um, 7.2 to 8.2\um, and
8.2 to 9.3\um , would have equivalent widths of about 4\um\ or $3.4\times
10^{13}$~Hz, 18\um\ or $9.2\times 10^{13}$~Hz, and 13\um\ or $4.9\times
10^{13}$~Hz, respectively (Helou \etal 2000).

The natural explanation for this continuum is a population of small grains
transiently heated by single photons to apparent temperatures near $1000$K.  
Such a population was invoked by Sellgren \etal (1984) to explain the 
3\um\  emission in reflection nebulae, and similar populations 
by other authors to explain the IRAS
12\um\ emission in the diffuse medium (e.g. Boulanger \etal 1988).
Small particles with ten to a
hundred atoms have sufficiently small heat capacities that a single UV
photon can easily propel them  to 1000~K equivalent temperature
(Draine \& Anderson 1985).  Such a population is a natural extension of the
AFE carriers, though it is not clear from these data whether it is truly
distinct, or whether the smooth continuum is simply the non-resonant
emission from the AFE carriers.  While the current data cannot rule out
other contributions to this continuum component, the shape does rule out a
simple 
extension of the photospheric emission from main sequence stars.  Red
supergiants and Asymptotic Giant Branch stars may contribute to the
continuum in this region but cannot dominate it, because their contribution
would not correlate with AFE from the ISM, and cannot match the shape of
a spectrum with $\rm f_\nu\propto \nu^{+0.65}$.

\subsubsection{High-Redshift Applications}

Since the spectral signature in Figure~4 applies to the majority of
star-forming galaxies, it can be used as a template to obtain redshifts of
highly extincted galaxies with the next infrared observatory, NASA's SIRTF
(Space InfraRed Telescope Facility).  For instance, a galaxy at a redshift
z=3 with a flux density average of 0.5 mJy in the range $19-27\um$ and a
total infrared luminosity comparable to Mkn~231 at $\sim 3\times
10^{12}${\lsun} would be detected by the SIRTF IRS (Infrared Spectrometer;
Roellig \etal 1998) in roughly 1000 seconds of integration (D.W. Weedman,
private communication).

In surveys with fixed spectral bands, the Aromatic Features will result in
a unique K-correction as redshift takes them in and out of the bands.  The
detection probability would be enhanced or reduced for certain redshift
intervals, causing ripples in source counts as a function of flux density; Xu \etal
(1998) have modelled this effect for NASA's WIRE (Wide-Field InfraRed
Explorer; Hacking \etal 1999) mission.  However, this effect has already
been manifested with ISO, in deep 15\um imaging data in the direction of
galaxy clusters.  These observations were aimed at detecting background
objects boosted by gravitational lensing, thus allowing the survey to
penetrate further in space and time.  In the Barvainis \etal (1999) survey,
one out of every five galaxies detected in the direction of each of
Abell~2218 and Abell~2219 turns out to be at a reshift near 1, the others
being at redshifts of 0.3 or less.  A redshift of 1 places the main 
AFE clump squarely in the 15\um band, enhancing the probability
of detection.  Similar results from a more extensive survey are reported by
Metcalfe \etal (1999), and by Fadda \etal (1999).

As the early Universe opens up for mid-infrared exploration, the properties
of low metallicity star-forming galaxies become more relevant as a template
for high-redshift galaxies.  ISO has provided the data for constructing
such templates, both from studies of low metallicity dwarf galaxies 
(Sauvage \& Thuan 1999), and the study of
nearby galaxy disks with significant metallicity gradients such as M~101
(Vigroux \etal 1999).

\subsubsection{Exceptions}

The ``standard'' spectra described above characterize the integrated
emission from normal galaxies, but exceptions arise at extreme conditions,
and are easiest to detect in specific ISM phases of individual galaxies.
At the smallest scales in the most intense parts of HII regions, Contursi
(1998), Tran (1998) and C\'esarsky (1996b) 
report spectra which depart significantly from the
standard ones in the location and strength of features.  Similarly
distorted spectra are also observed in NGC~4418 (Lu \etal 2000), and close
to the Active Nucleus of Cen~A (Vigroux \etal 1999).  This
radical transformation in the spectrum must then result from extremely
UV-rich heating radiation, and most likely reflects severe modification of
the emitting grains, including the destruction of Aromatic Feature
carriers.  Significant distortions in the spectrum driven by a rising
continuum but without radical changes in the Aromatic Features have also
been reported by Lutz \etal (1998) in extremely active galaxies.  The
mid-infrared spectra of starbursts are relatively ``standard'', whereas
galaxies with an active Black Hole nucleus (AGN) have a continuum rising towards
longer wavelengths with insignificant AFE.  Ultra-Luminous
Infrared Galaxies show intermediate spectra, modified in addition by large
optical depths (Genzel \etal 1998; Laurent \etal 1999).

At the other end of the heating sequence, C\'esarsky \etal (1999)
report that the 5--9\um AFE are missing from the
emission of the bulge and of the quiet parts of the bright infrared ring in
M~31.  The 11.3 and 12.7\um features are present, perhaps a result of their
originating primarily on neutral PAHs.  The relatively weak and UV-poor
heating radiation in these parts of M~31 cannot alone explain the absence
of Aromatic Features, for other studies (e.g. Uchida \etal 1998, Uchida \etal\ 
1999) show
that photons of a few eV are sufficient to generate the standard spectrum.
It would appear that the dust properties in M31 are again modified, this time by
prolonged shielding from UV processing in the astration cycle.  Additional
evidence of abnormal ISM properties in the same regions are found by
Pagani \etal (2000) in their study of the broad-band mid-infrared
emission.

Another peculiar exception is provided by Sauvage \& Thuan (1999)
in SBS~0335-052, a blue compact dwarf with a metallicity about
1/40 solar.  No AFE are detected here, and the
spectrum appears best fit by highly extincted blackbody emission.  The
authors propose the early chemical age of the system as the most likely
reason behind the absence of the AFE, though one cannot rule
out the possibility that the UV radiation from the intense star formation
episode also plays a role in the destruction of the Aromatic Feature
carriers.  If this spectrum does indeed characterize all low metallicity
objects, one would expect the ``standard'' spectrum to disappear gradually
as a function of increasing redshift, thus making it harder to detect the
Aromatic signature of dust and distinguish star formation from Black Hole
activity as the energy source in the earliest systems.

\subsection {The ISO-IRAS Color Diagram}

A large number of extragalactic broad-band measurements was collected by
ISOCAM at wavelengths between 4 and 18\um, most frequently using the 
``LW2'' band centered at 6.75\um, and the ``LW3'' band centered at 15\um.  
The LW2 filter was
designed to capture mostly AFE emission, and LW3 was aimed at the
continuum range beyond the bulk of the AFE, though it ends up with a small
contribution from the 12.5\um feature.  
The 6.75-to-15\um\ color ratio has 
emerged as an interesting diagnostic of the radiation environment.  It
remains relatively constant and near unity as the ISM of galaxies
proceeds from quiescent to mildly active.  As dust heating increases
further, the 15$\um$ flux increases steeply compared to 6.75$\um$, pointing
to a significant contribution by dust at color temperature
$\rm100~K<T_{MIR}<200~K$, typical of a heating intensity up to $10^4$ times
that of the diffuse interstellar radiation field in the local Milky Way
(Figure~6; Helou \etal 1997; Helou 1999).  
While such a temperature could result from classical
dust heated within or just outside HII regions, there is no decisive
evidence as to the size of grains involved.  It is simpler at this time
to associate this component empirically with the observed emission
spectrum of HII regions and their immediate surroundings (Tran 1998;
Contursi 1998).  This emission has severely depressed AFE, and is dominated
by a steeply rising though not quite a blackbody continuum, consistent with
mild fluctuations in grain temperatures, $\Delta{\rm T}/{\rm T}\sim 0.5$.
This HII region hot dust component at $\rm T_{MIR}$\ 
becomes detectable in systems where the
color temperature from the 60-to-100\um\ ratio is only $\rm T_{MIR}/2$,
demonstrating the broad distribution of dust temperatures within any
galaxy.  The combined data from ISO and IRAS on these systems are
consistent with an extension of the ``two-component model'' of infrared
emission (see \S 3.4 above) and demonstrate the fallacy of modeling the infrared
spectra of galaxies as single
temperature dust emission.  The low 6.75-to-15\um\ color ratio is 
associated with the active component, and combines in a variable 
proportion with a component with a 6.75-to-15\um\ near unity.  
This color behavior was observed
in the sample of galaxies used for the Key Project on normal galaxies
(Helou \etal 1996, Silbermann \etal 1999), and confirmed
in the sample of galaxies observed under ISOCAM Guaranteed Time 
(see Figure 1 in Vigroux \etal 1999).  See also \S 4.4 below.

\subsubsection{The Global Infrared Spectrum}

Turning now to the long wavelength end of the spectrum,
photometry at $120-200\um$ using ISO-PHOT is
starting to constrain the distribution of dust temperatures at low heating
levels, especially in nearby well resolved galaxies such
as M31 (Haas \etal 1998, 1999), 
M51 or M101 (Hippelein \etal 1996),  where cold dust
dominates the luminosity.  Similar analysis on more active galaxies is also
under way ({\it e.g.} Klaas \etal 1997; Klaas, Haas \& Schulz 1999) 
to obtain the best possible estimates 
of the total infrared emission and therefore of the dust mass.  Alton \etal
(1998) have reported that the emission
is more extended at 200\um\ than 
at shorter wavelengths in several galaxies.  This result
however hinges closely on a very precise knowledge of the beam shapes at 
various wavelengths, which was not yet achieved at the time of that publication.

A proper determination of the total infrared luminosity and the 
long-wavelength spectral shape in normal galaxies is critical to 
estimating  the contribution of galaxies to the infrared and submm 
extragalactic background light, and thereby deriving the infrared term 
in the star formation history of the Universe.  While work
on the ISO-PHOT calibration continues, one could estimate an improved 
infrared spectral energy distribution by combining the mid-infrared results
described above with existing IRAS data (Figure~7).  
Such a rough estimation is presented
in Table~1 for several different levels of activity in galaxies, parametrized
by the 60-to-100\um ratio.  It should be noted that all information 
at $\lambda>100\um$ in Table~1 is based on modeling IRAS data using a power-law 
distribution of dust mass as a function of heating intensity, and does not
use any empirical constraint (Dale \etal 2000a; Helou \etal 2000).
Table~1 illustrates the general trend, but   also ignores  variations in 
spectral shape at constant 60-to-100\um\  ratio, 
including intrinsic scatter in the ratio of mid-infrared to far-infrared 
(Lu \etal 1999), and dispersion in the 25-to-60\um ratio.  
The 20 to 40\um range appears to show the most significant
growth in fractional terms at the expense of the submillimeter 
as the activity level increases, suggesting that the $20-40\um$\  continuum 
may be the best dust emission tracer of current star formation in galaxies.  
Even after ISO, our knowledge of the detailed
shape of the spectrum at $\lambda>100\um$\  remains model-dependent, and may not 
improve significantly until the launch of SIRTF, and then of FIRST (Far infraRed
and Submillimeter Telescope).

\begin{table*}[htb]
  \caption{\em Rough energy distribution across the infrared spectrum
    for galaxies with various levels of star formation activity,
    parametrized by the IRAS color in the first column.  Each column 
    heading gives the wavelength range over which the spectrum
    is integrated, and the table entries are the fractions of total
    infrared luminosity appearing in that range.  The spectral
    range in column 6 corresponds to the ``FIR'' synthetic
    band (Helou \etal 1988).}
%{{\F3}\over{\F4}}
  \label{tab:table*}
  \begin{center}
    \leavevmode
    \footnotesize
    \begin{tabular}[h]{ccccccc}
      \hline \\[-5pt]
{${\rm f_\nu (60\um)}\over{\rm f_\nu (100\um)}$} & F(3--5\mic) & F(5--13\mic) & F(13--20\mic) & F(20--42\mic) & F(42--122\mic) & F(122--1100\mic) \\[+5pt]
      \hline \\[-5pt]
      0.40    & 0.024 & 0.122 & 0.033 & 0.10 & 0.41 & 0.32 \\
      0.63    & 0.017 & 0.086 & 0.037 & 0.18 & 0.50 & 0.19 \\
      1.00    & 0.008 & 0.048 & 0.043 & 0.28 & 0.54 & 0.09 \\[+5pt]
      \hline \\
      \end{tabular}
  \end{center}
\end{table*}

\subsection {A Mid-Infrared Look Within Galaxies}

ISO-CAM CVF studies between 5 and 17\um\ are turning out to be powerful
diagnostics of the radiation field within the disks of nearby galaxies,
allowing us to disentangle the variations in heating intensity and hardness
of interstellar radiation.  The approach is to relate the intensity to the
shape of the continuum, and the hardness to the ratios of ionic
fine-structure lines (Tran 1998; Contursi 1998).  
See also the overview on ISOCAM studies of
nearby galaxies by Vigroux \etal 1999, and the studies of NGC~891 by Le
Coupanec \etal (1999) and by Mattila \etal (1999).  Such studies are valuable 
in establishing the local relation between mid-infrared emission and the
star formation intensity, thereby guiding the interpretation of the
global fluxes.

The ISOCAM images of galaxies show dust emission in nuclear regions, in the
inner barred disk, outlining the spiral arms, and tracing the disk out to
the Holmberg radius and beyond (Malhotra \etal 1996, Sauvage \etal 1996,
Vigroux 1997, Smith 1998, Roussel \etal 1999, Dale \etal 2000b).  
There are clear color variations within spiral galaxies, some
of which have not yet found satisfactory explanations (Helou \etal 1996;
Tran 1998; Vigroux \etal 1999).  Dale \etal (1999) describe 
behavior similar to the ISO-IRAS color diagram within the 
disks of three star forming galaxies, IC~10, NGC~1313, and NGC~6946,
where the 6.75-to-15\um\ color drops precipitously as the surface
brightness exceeds a certain threshold.  The point of
inflexion in the color curve occurs at a surface brightness which
is a function of the dust column density, whereas the shape of the
curve seems invariant, and may result from a rise in both the hardness and
intensity of the heating radiation (Figure~8).  Dale \etal discuss these findings
in the context of a two-component model for the interstellar medium,
suggesting that star formation intensity largely determines the 
mid-infrared surface brightness and colors within normal galaxy disks,
whereas differences in dust column density  are the primary drivers of 
mid-infrared surface brightness variations among galaxy disks.

Rouan \etal (1996), Block \etal (1997) and Smith (1998) have combined
ISOCAM and Br$\gamma$ images with other broad-band and line images to
estimate star formation rates, ISM parameters, obscuration and dust
properties.  These studies again point to AFE carriers as a
ubiquitous component of interstellar dust, to the likely destruction of
these carriers by ionizing UV, and to dust heating in non-starburst disk
galaxies being derived from both old stars and OB stars.

In M31, Pagani \etal (2000) demonstrate a very close correlation 
between mid-infrared emission at both 6.75 and 15\um and the distribution of 
neutral gas as traced by HI and CO maps.  The correlation is poorer with
ionized gas as traced by H$\alpha$, and poorest with UV emission, a result which
they attribute to extinction.  They conclude that AFE can be
excited by visible and near-IR photons, the dominant dust heating
vectors in this particular case, and therefore by
older disk and bulge stars. They also find evidence that in this environment
the AFE carriers are
amorphous carbonaceous particles formed in the envelopes of carbon stars, 
and have not yet been graphitized by ultraviolet radiation.

\section{A WALK IN THE LINE FOREST}

The unfettered access to the infrared spectrum in space gave ISO a tremendous 
advantage in studying any infrared line in galaxies regardless of redshift,
until the lines leave the window at high redshifts.  The low thermal 
backgrounds of a cryogenic telescope in space allowed it 
to tackle much lower surface brightness sources.  As a result, 
ISO extended the spectroscopic studies 
of starburst nuclei that the Kuiper Airborne Observatory (KAO) had carried 
out to the less intense star formation in normal galaxies.
The spectrocopic 
capabilities of all four instruments were especially valuable tools 
in characterizing the interstellar gas and radiation field, 
and in constraining the overall energetics and star formation rate.  

Many lines in the infrared range carry substantially more luminosity 
than the lines most studied from the ground and normally used for probing 
the non-ionized ISM.   The HI~$\lambda$21~cm  fine-structure line,
popular because its flux can be safely assumed to be proportional to 
the total emitting HI mass, carries about $10^{-9}$~\lfir.
The CO pure rotational lines in the millimeter and submillimeter range 
(J=1$\rightarrow$0, 2$\rightarrow$1, 3$\rightarrow$2, etc. at 2.3~mm, 
1.15~mm, etc.) carry a few times $10^{-6}$~\lfir.
By contrast, the molecular hydrogen rotational lines discussed below
carry a  few times $10^{-4}$~\lfir, and several of the infrared 
fine-structure lines such as $\lbrack$CII$\rbrack$,  
$\lbrack$OI$\rbrack$, $\lbrack$SiII$\rbrack$,  
or $\lbrack$OIII$\rbrack$ carry $10^{-3}$\  to  $10^{-2}$~\lfir.
They are thus by far more significant as a measure of the energetics
of interstellar gas, and easy to detect at greater distances, making
them valuable tools for probing the star formation process in the 
most distant, youngest galaxies.

\subsection {Molecular Lines}

One of the great achievements of ISO was the detection of the long-sought
molecular hydrogen transitions from rotationally excited states, and the
realization of their promise as accurate tracers of the H$_2$ mass ({\it
e.g.}  Draine \& Bertoldi 1999), allowing for the first time a direct gauge 
of the dominant phase of the star-forming ISM.  Most of the H$_2$ data
were collected by the SWS (de Graauw \etal 1996), and their interpretation
and analysis in the Milky Way is discussed by
Cox \& Boulanger elsewhere in this volume.  
Several of these lines,
especially S(0), S(1) and S(2) were detected in galaxies early in the ISO
mission, permitting a detailed discussion of the gas phase conditions and
heating mechanisms, as in the study by  Valentijn \etal (1996)  of the
nuclear star burst of NGC~6946.  Because of their high excitation levels
however, these lines tend to trace H$_2$ warmer than $\rm \sim 100~K$, and
are therefore easier to detect in the more intense star formation
environments (Kunze \etal 1996; Kunze \etal 1999).  Still,
Valentijn \etal (1999) report the detection of H$_2$
emission from the extended disk of the edge-on galaxy NGC~891.  They detect
S(0) and S(1) at eight positions, tracing the emission out to 12~kpc from
the nucleus of the galaxy, and derive H$_2$ temperature constraints and
molecular mass estimates.

Hydroxyl (OH) was also detected in starburst galaxies, but in absorption
rather than emission.  A first report by Skinner \etal (1997) showed OH
absorption at 35\um in the spectrum of Arp~220, and confirmed for the first
time pumping by infrared photons as the excitation mechanism behind OH
mega-masers.  Bradford \etal (1999) report OH in absorption at
35, 53 and 119\mic~ in NGC~253, and resolve the line at 119\mic~ with the
LWS Fabry-P\'erot mode.  They estimate total column and excitation
temperatures for the OH, and constrain the geometry of the molecular
material and its relationship to the infrared-emitting dust.  

\subsection {Fine-Structure Lines}

Far-infrared fine structure lines, $\lbrack$CII$\rbrack$~$\lambda
157.7\um$\ and $\lbrack$OI$\rbrack$ ~$\lambda 63.2 \um$ in particular, have
long been used for estimating density and radiation intensity in
photo-dissociation regions (PDR) ({\it e.g.}  Hollenbach \& Tielens 1997),
which are the interfaces between HII regions and molecular clouds.
These dense ($\rm {\it n}\sim10-10^5cm^{-3}$ or more), 
warm ($\rm T\sim 100-300~K$), 
neutral media are preferentially cooled by $\lbrack$CII$\rbrack$ 
because carbon is abundant,
easy to ionize (IP=11.26~eV), and easy to excite ($\rm \Delta E/k\sim90~K$).
At $\rm T\ge 200~K$ and $\rm {\it n}\ge10^5cm^{-3}$, $\lbrack$OI$\rbrack$ 
takes over as the main coolant, with its higher excitation threshold
($\rm \Delta E/k\sim224~K$).  In PDRs, both transitions are excited predominantly
by electrons which have been extracted from dust particles by ultraviolet
photons usually assumed to have $\rm \ge6eV$.  This well known and studied
photo-electric effect (Field, Goldsmith \& Habing 1969;
Hollenbach \& Tielens 1997) is estimated to
have a yield $\le~10^{-2}$ in an ISM illuminated by starlight.
Traditionally, various line ratios and line-to-continuum ratios are
used to constrain the main parameters of PDR regions, by comparison
to calculations from models of slab PDRs.  In such models, the emission 
is integrated along a line of sight sampling the PDR, starting at the HII 
region front, through the molecular cloud, and on towards the molecular 
cloud core, to the point where emission becomes negligible.

Just as $\lbrack$CII$\rbrack$ and $\lbrack$OI$\rbrack$ are important coolants
of the neutral ISM, so are $\lbrack$NII$\rbrack$~$\lambda 121.9\um$, 
$\lbrack$OIII$\rbrack$~$\lambda 88.4\um$\ and $\lambda 51.8\um$, and
$\lbrack$NIII$\rbrack$~$\lambda 57.3\um$\ significant for HII regions.
While ISO collected substantial data on these lines, they will not be 
discussed in any detail in this review because little has been published
yet on this topic.  It should be kept in mind however that $\lbrack$CII$\rbrack$ 
can arise in HII regions as well as PDRs, and that complicates the 
interpretation of line ratios.  Similarly, the continuum emission from
dust will arise from a variety of media, complicating the interpretation
of line-to-continuum ratios (\S 5.3).

ISO-LWS (Clegg \etal 1996) has provided a wealth of data,
whose interpretation is creating controversy and challenging theoretical
models.  Using the Normal Galaxy Key Project sample, 
Malhotra \etal (1997) showed that while two thirds of normal
galaxies have $\cal L{\rm(CII)}/\cal L\rm{(FIR)}$ in the range $2-7\times
10^{-3}$, this ratio decreases on average as the 60-to-100\um\ or the $\cal
L{\rm(FIR)}/\cal L\rm{(B)}$ ratios increase, both indicating more active
star formation (Figure~9).  The same CII deficiency is also observed in 
ultra-luminous infrared galaxies such as Arp~220 (Luhman \etal 1998).
Malhotra \etal\  linked this decrease to elevated heating
intensities, which ionize grains and thereby reduce the photo-electric
yield.  They discussed other possible causes, such as self-absorption of [CII],
heating by non-ionizing stars, or the influence of an AGN.  Optical depth
effects were dismissed because no deficiency trend is observed in [OI]
even though it is expected to have greater optical depth than [CII].
The possibility of non-ionizing stars was dismissed since it required the most
unlikely scenario that such stars would dominate the heating systematically
in the most actively star forming galaxies, including objects such as
Arp~220 (Fischer \etal 1999).  Finally, since the normal galaxy sample was
selected to avoid AGN, the latter are unlikely to be the sole reason behind
the [CII] deficiency trend.  Fischer \etal (1999)
and Malhotra \etal (1999) 
give updated discussions of this topic, while Lord \etal (1999) and Unger \etal
(1999) discuss PDR properties in NGC~4945 and Cen~A.

%  where optical depth effects have been proposed as the origin of the effect.  
%  This is hard to reconcile however
%  with a similar deficiency effect observed for 
%  $\lbrack$NII$\rbrack$ ~$\lambda 121.9 \um$, but not for 
%  $\lbrack$OI$\rbrack$ ~$\lambda 63.2\um$ (Figure~10).

The relation of the $\rm\lbrack CII\rbrack$ line luminosity to the total
star formation rate in a galaxy has been debated just as vigorously as
other infrared observables (Stacey 1991).  One of the outstanding questions
there has been the importance of low density HII regions
as a source of $\rm\lbrack CII\rbrack$ emission.  Stacey \etal (1999) 
present detailed maps of M~83 in several fine-structure
lines, and address this question directly, estimating that 27\% of the
total emission may well originate in that diffuse component.
This same contribution
by diffuse PDRs is invoked by Pierini \etal (1999) in explaining
the high $\rm\lbrack CII\rbrack$-to-CO ratios in quiescent Virgo cluster
galaxies.  
Using the same Virgo galaxies data, Leech \etal (1999)
report a trend of decreasing $\rm\lbrack CII\rbrack$/FIR ratios as galaxies
become less active in star formation.  Combining the results of Leech \etal 
with those of Malhotra \etal  leads to a picture where 
$\rm\lbrack CII\rbrack$/FIR rises by an order of magnitude as galaxies
move away from complete quiescence,
reaches a broad maximum for normal galaxies actively forming stars,
then decreases again by more than an order of magnitude 
as galaxies begin to approach the extreme properties of starbursts.

\subsection {Interpreting the PDR Lines}

What could be behind this behavior?  At the quiescent end, the 
relative lack of ionizing photons in the stellar spectrum could 
naturally explain the low values of $\rm\lbrack CII\rbrack$/FIR.
UV-poor heating would on one hand lower CII abundance,
and on the other hand yield less energetic photo-electrons, 
so a less energetic CII excitation would be expected.
At the starburst end, the extreme excitation conditions could generate
lower photo-electric efficiency as proposed by Malhotra \etal (1998),
and as detailed in the previous section.

While that picture is physically reasonable, there is evidence however 
favoring a different interpretation at least at the high excitation end
of the sequence, namely that the relative drop in $\rm\lbrack CII\rbrack$ 
is due to a decrease in the concentration of the grains crucial for
the photo-electric effect.  It is well known 
({\it e.g.} Hollenbach \& Tielens 1997) that the smallest grains such
as the carriers of AFE are responsible for the bulk of photo-electric
yield in a PDR.  Depletion of these carriers would decrease
the coupling between radiation field and gas, and lead to weaker
fine-structure line emission compared to total dust re-radiation.  
The evidence in question is that unlike
$\rm\lbrack CII\rbrack$/FIR,  the ratio
of $\rm \lbrack CII\rbrack$ to AFE flux does
not decrease in normal galaxies with increasing activity (Helou \etal 2000).
This suggests that the emissions from $\rm\lbrack CII\rbrack$ and AFE 
originate in the same regions of the ISM, so their ratio is dictated
by the physics of the photo-electric effect and remains constant. 
On the other hand, $\rm\lbrack OI\rbrack$
and FIR originate from distinctly warmer and denser regions, so that the 
changing ratio $\rm\lbrack CII\rbrack$/FIR, just like 
$\rm\lbrack CII\rbrack$/$\rm\lbrack OI\rbrack$, reflects the 
systematic shift in density distribution of the ISM as galaxies 
approach star-burst conditions.  

At the high end of activity, the empirical evidence for decreased 
AFE carriers is quite clear (Helou, Ryter \& Soifer 1991; Genzel \etal 1998), 
and the evidence is strong that they are 
destroyed by  the intense UV radiation from massive stars.  
There is less empirical evidence however for a systematic lack of 
Aromatic grains at the low end of the excitation sequence.  Perhaps
the most interesting hints come from the detailed investigation of 
the mid-infrared emission by C\'esarsky \etal (1999) in M31. 
It is still a matter of debate whether AFE carriers condense out on
larger grains and are re-extracted by exposure to UV radiation (Boulanger
\etal 1990), or whether the AFE carriers emerge only as a result of the
first photo-processing by UV light of newly formed amorphous carbon grains
(Pagani \etal 2000).

Most of the interpretation of [CII] and [OI] line fluxes is done in the
context of PDR theories and models, which provide the most powerful tools
available.  It should be recalled however that this context assumes
implicitly that all  observed fluxes in line and continuum originate in
PDRs.  The broadest definition of PDR, usually supported in model calculations,
includes any non-ionized region heated by stellar photons.  Thus defined,
PDRs do provide all [OI] emission from galaxies, but  not all [CII] or 
dust continuum.  Indeed, both of the latter can have contributions from
HII regions of all descriptions, from the compact to the diffuse, extended
low density variety.  Additional flux of either [CII] or continuum
will throw off the ratios based on which PDR models derive density
and heating intensity.  One can mitigate the problem by estimating 
HII regions contributions based on the [NII] line for instance, and
adjust the [CII] flux accordingly (Malhotra \etal 2000.  See also \S 7.1).

\section{MORE STUDIES}

In addition to the topics covered above, the ISO archive will soon offer a
wealth of data on a wide variety of extragalactic subjects affecting our
understanding of normal galaxies.  Studies in galaxy clusters and groups
for instance have targeted the population of medium-distant clusters at
z=0.1--0.3 (L\'emonon \etal\  1999, Fadda \& Elbaz 1999;
Metcalfe \etal\  1999).  
Stickel \etal (1998) have reported the detection of
intracluster dust in Coma at 200\mic, and related it to the cooling flows
revealed by X-ray observations, while Xu, Sulentic \& Tuffs (1999) have
studied intergalactic star formation in Stefan's Quintet using mid-infrared
data.  Charmandaris \etal (1999) and Gallais \etal (1999) have taken a
detailed look at some of the nearest interacting galaxy systems in order to
study collisionally induced star formation.

Beichman \etal (1999) and Gilmore \& Unavane (1998)
report on deep searches for mid-infrared haloes around
edge-on galaxies, and place upper limits on the contribution to the
missing mass by low-mass stars.  Jarrett \etal (1999) 
report on mid-infrared detection of a bridge of interstellar gas and
magnetized plasma created between two galaxies after a collision which
caused the two disks to pass through each other and entangle
the interstellar media.

Observations of Local Group members allow detailed analysis of
the ISM and star formation on the few parsec scale, and are a crucial 
stepping stone to understanding the integrated emission in more
distant objects.  Such detailed studies of
HII and star formation  regions, or of the stellar population
in the Magellanic Clouds are reported by 
Comer\'on \& Claes (1999), Henning \etal (1999), Loup \etal (1999), 
Missoulis \etal (1999), and 
Vermeij, van der Hulst \& Baluteau (1999).

\section {TOMORROW'S INFRARED GALAXIES}

\subsection {Challenges and Prospects}

The two-component model approach is clearly an over-simplification of the
complexities in a real system, representing what is almost certainly a 
continuous distribution over some parameter by a linear combination of
the extremes.  One challenge ahead is to formulate more realistic descriptions 
of star-forming galaxies, incorporating a distribution of ISM mass over
the key parameters density, intensity and hardness of heating radiation.
Additional parameters such as cosmic ray flux, metallicity, dust abundance 
and properties, or X-ray flux might be added as secondary parameters.
From such a description, one would estimate observables as integrals
over the phase space defined by the key parameters (see \S 2 above), 
and translate empirical
properties of galaxies into constraints on the ISM properties.

A related challenge is to formulate a realistic description of the geometry
of dust, gas and heating stars.  The ISM appears clumpy on all scales,
perhaps best described as fractal in its structure 
as evident from IRAS, HI or CO maps of emission (Falgarone \& Phillips 1996).
Detailed studies of nearby PDRs suggest substantial ultraviolet penetration
far from the boundary of the HII region (Howe \etal 1991; see also discussion
in \S~2.1 of Hollenbach \& Tielens 1997).
Geometry may well be a critical factor in some of the observed properties 
of galaxies; it may well affect the averaging over regions and directions
in ways that inject apparent simplicity in otherwise quite complex systems.
Detailed models may be the only way to establish the role of geometry 
in normal galaxies.

Neglecting for now the complexities above, one can expect substantial progress
in understanding normal galaxies by exploiting ISO data with new approaches.
In addition to the powerful spectroscopic techniques offered by ISO,
the improved spatial resolution of ISO-CAM should allow us to go beyond
luminosity, and work with surface brightness in the infrared.  This is a
distance-independent quantity which can be measured and compared in nearby
as well as distant  galaxies, and combines information on the intensity 
of heating radiation and the column density of dust, and therefore might
help us disentangle these two quantities using additional information from
other observables.  Some work along these
lines based on IRAS data has been published (Wang \& Helou 1994; 
Helou \& Wang 1995; Meurer \etal 1997),
but ISO data offer much better  sensitivity and spatial resolution 
(Dale \etal\ 1999; Dale \etal\  2000a).

With ISO data in hand, it should be possible to answer some basic questions
about galaxies, deriving for instance an accurate star formation rate from
a combination of observables, and attaching specific physical meaning to 
standard paramaters commonly used today, such as IRAS colors,
line-to-continuum ratios, or infrared-to-blue ratios.  More difficult
but probably within reach of ISO data interpretation would be a means
to distinguish between physical parameters and mixing ratios, so one
can tell whether two galaxies differ because their PDRs are systematically
different, or because they have a different mix of PDR, HII regions and
molecular clouds.

\subsection {Suggestions}

The study of normal galaxies will remain exciting because of, and in spite of,
the complexities and bewildering abundance of data and correlations.  
A few suggestions might
help in navigating these complexities:  

\begin{enumerate}

\item In statistical studies,
it is critical to understand the sample being used, its biases
and limitations.

\item The questions to pursue should be physical rather than statistical
in nature.  The latter will flow from the former, but should not overshadow them.

\item Quantities studied, statistically or in detail, should have
clear physical significance.  Distance-independent ``intensive'' quantities are
preferable to luminosities or similarly scaling extensive parameters.

\item In constructing quantities from observables, one should avoid 
complex parameters with multiple built-in assumptions, such as dust mass
which combines a flux with an uncertain color temperature taken to a power
of 5 or so.
\end{enumerate}

\section {CONCLUSION}

Normal galaxy studies with ISO have already yielded major progress beyond 
the knowledge of the IRAS era, improving dramatically our empirical understanding 
of mid-infrared spectra, mid-infrared images, the global spectral energy
distribution and the far-infrared fine-structure lines.  This progress has already 
led to rethinking plans for other surveys and missions addressing those observables.

ISO data are also re-shaping our physical understanding of galaxies:
allowing for a better decomposition of the infrared luminosity from a
combination of spectral energy distribution and imaging;  providing narrower
constraints on density and radiation intensity from analysis of 
fine-structure lines, surface brightness, and the spectral energy distribution; 
providing a broader perspective on objects with extreme values of 
luminosity or star formation rate, based on an understanding of the
sequence that lies in-between these extremes.
Much remains to be learned from ISO, suggesting several
years of productive research on the ISO
Science Archive still lie ahead. 

SIRTF (Space InfraRed Telescope Facility), scheduled for a December~1, 2001 
launch, will build on ISO and tackle the new puzzles with its greater sensitivity,
finer spatial resolution and larger fields of view.  SOFIA 
(Stratospheric Observatory for Infrared Astronomy) is expected to
fly starting in 2002,  bringing to the field greater spatial and 
spectral resolution and instrumental versatility.  FIRST 
(Far-Infrared and Submillimeter Telescope) will then open
up the submillimeter spectrum from space starting in 2007, providing access
to the colder material in molecular clouds, so all   ISM phases from
cold clouds to HII regions and shocked gas can be studied, bringing into
focus the full picture of the star formation cycle.

\acknowledgments

I would like to thank the organizers of the Summer School for the invitation
to participate in this exciting encounter, and for their patience when it came
to submitting the manuscript.  James Lequeux in addition provided help and
comments on the manuscript.  
Danny Dale and Alessandra Contursi helped with 
customized plots, and a careful reading of the manuscript.  
This work was carried out at the California Institute of Technology, under
funding by the National Aeronautics and Space
Administration. It was supported in part by the ISO Data Analysis Funding
Program, administered for NASA by the Jet Propulsion Laboratory, 
California Institute of Technology.

\newpage

%%%%%%%%%%%%%%%%%%%  FIGURE 1
 \begin{figure*}[b]
%  \begin{center}
%    \leavevmode
  \centerline{\psfig{file=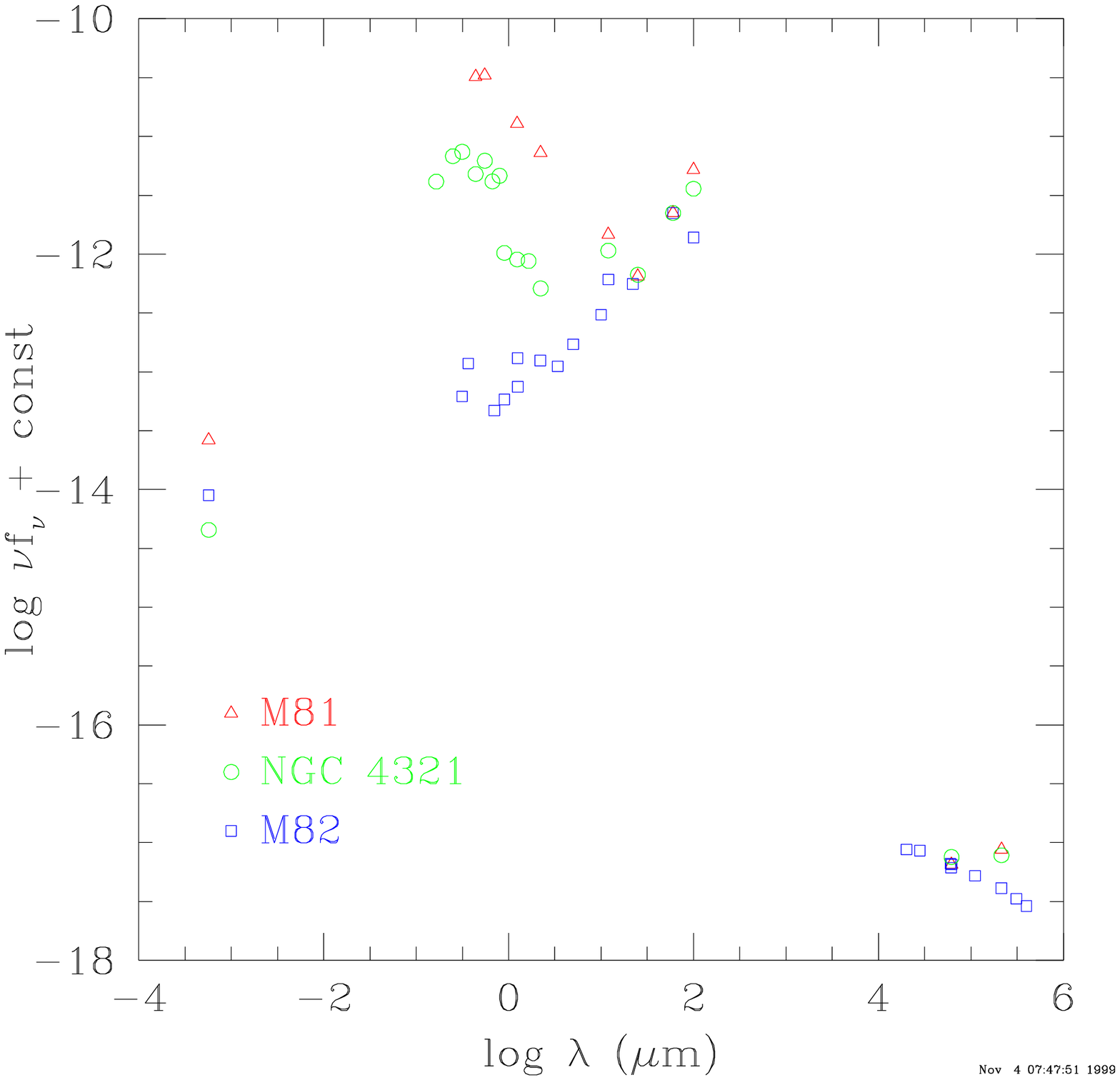,width=18.0cm}}
%  \centerline{\bf\it fig01.ps}
%  \vspace{5.8cm}
%  \end{center}
  \caption{\em The full spectral energy distribution for three representative
   star-forming galaxies.  Note in particular the variation in the relative
   importance of the infrared and the visible-ultraviolet bands.
   Spectra are normalized to the same ordinate at 60$\mu$m.
   }
  \label{fig:fig-1}
\end{figure*}

%%%%%%%%%%%%%%%%%%%  FIGURE 2
 \begin{figure*}[b]
%  \begin{center}
%    \leavevmode
  \centerline{\psfig{file=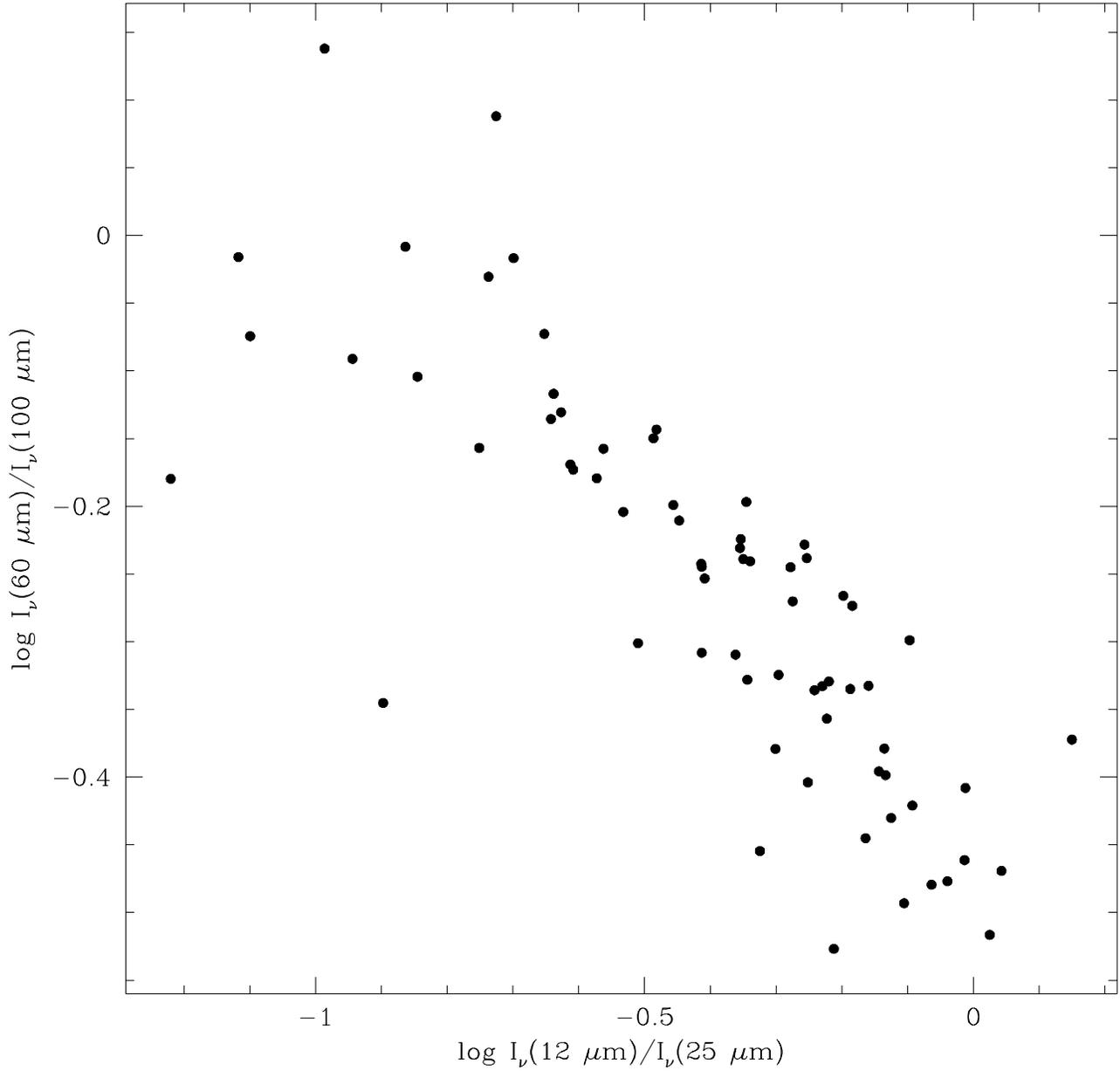,width=18.0cm}}
%  \centerline{\bf\it fig02.ps}
%  \vspace{5.8cm}
%  \end{center}
  \caption{\em The IRAS color-color diagram describes the variation in the
   shape of the infrared emission from interstellar dust as the star formation
   activity and dust heating vary in the galaxy.  This plot shows the 
   data for the sample used in the ISO Key Project on Normal Galaxies 
   (cf. \S 4.1).
   }
  \label{fig:fig-2}
\end{figure*}

%%%%%%%%%%%%%%%%%%%  FIGURE 3
 \begin{figure*}[b]
%  \begin{center}
%    \leavevmode
  \centerline{\psfig{file=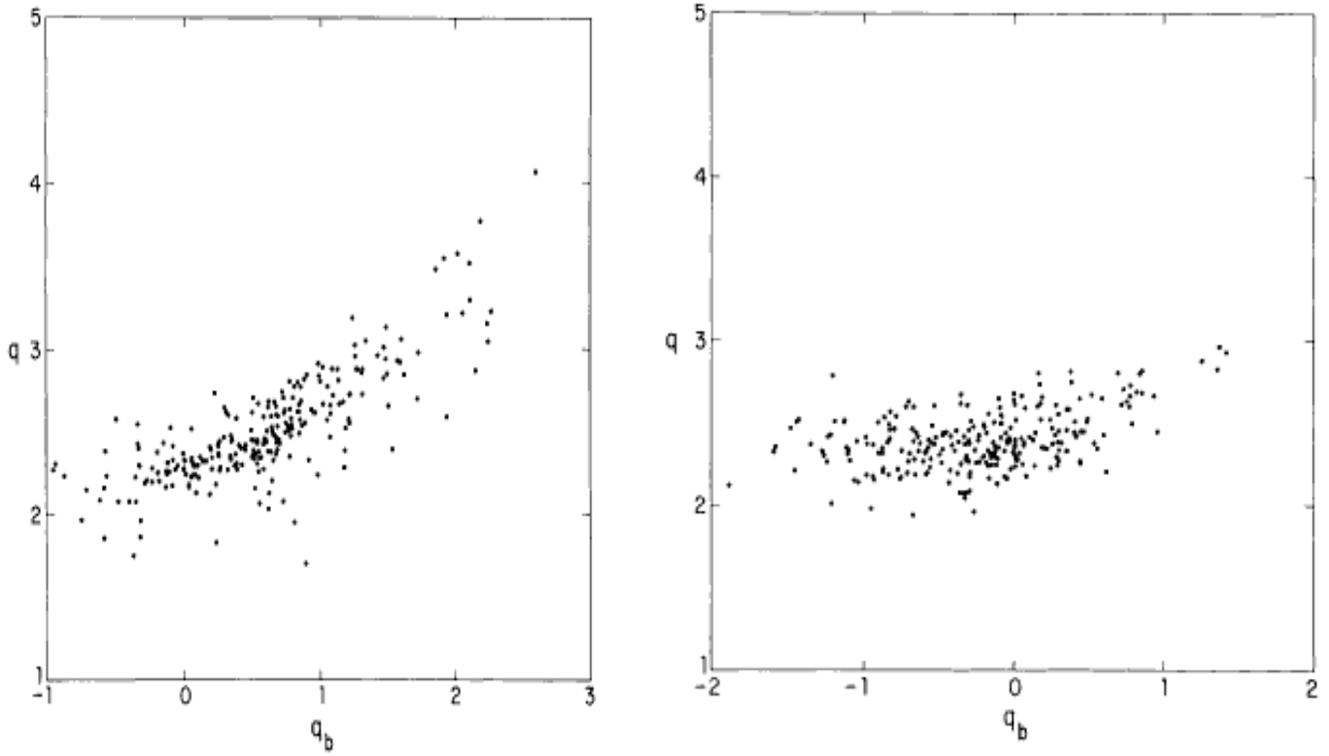,width=18.0cm}}
%  \centerline{\bf\it fig03.ps}
%  \vspace{5.8cm}
%  \end{center}
  \caption{\em The strong relation between infrared and radio is best shown
   in this figure from Condon, Anderson \& Helou (1991), clearly illustrating
   the low dispersion in the ratio $Q$ in the more active galaxies.
   The left-hand-side frame shows data from the IRAS Bright Galaxy Sample
   (Soifer \etal 1986), an infrared-selected sample with a preponderance
   of galaxies dominated by on-going star formation.  The right-hand-side
   frame shows data for a sample drawn from the Revised Shapley-Ames catalog,
   whose visible-light selection gives a more quiescent galaxy population.
   }
  \label{fig:fig-3}
\end{figure*}

%%%%%%%%%%%%%%%%%%%  FIGURE 4
 \begin{figure*}[b]
%  \begin{center}
%    \leavevmode
  \centerline{\psfig{file=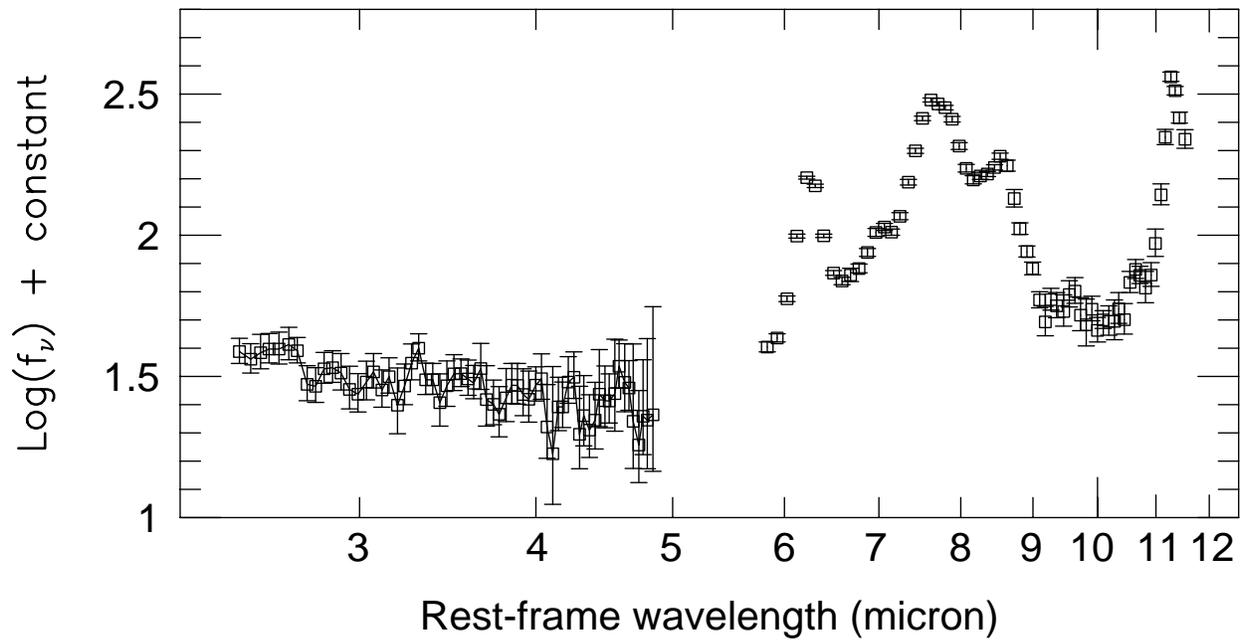,width=16.5cm,bbllx=64pt,bblly=362pt,bburx=523pt,bbury=604pt}}
%  \centerline{\bf\it fig04.ps}
%  \vspace{5.8cm}
%  \end{center}
  \caption{\em  A composite mid-infrared spectrum obtained from a straight
   average of a set of PHT-S spectra of 28 galaxies (Helou \etal 1999).
   Error bars indicate the dispersion among the avaraged spectra 
   when they are all normalized to the flux integral between 6 and 6.6\um.
   Note the ordinate is the flux density per frequency interval.
   }
  \label{fig:fig-4}
\end{figure*}

%%%%%%%%%%%%%%%%%%%  FIGURE 5
 \begin{figure*}[b]
%  \begin{center}
%    \leavevmode
  \centerline{\psfig{file=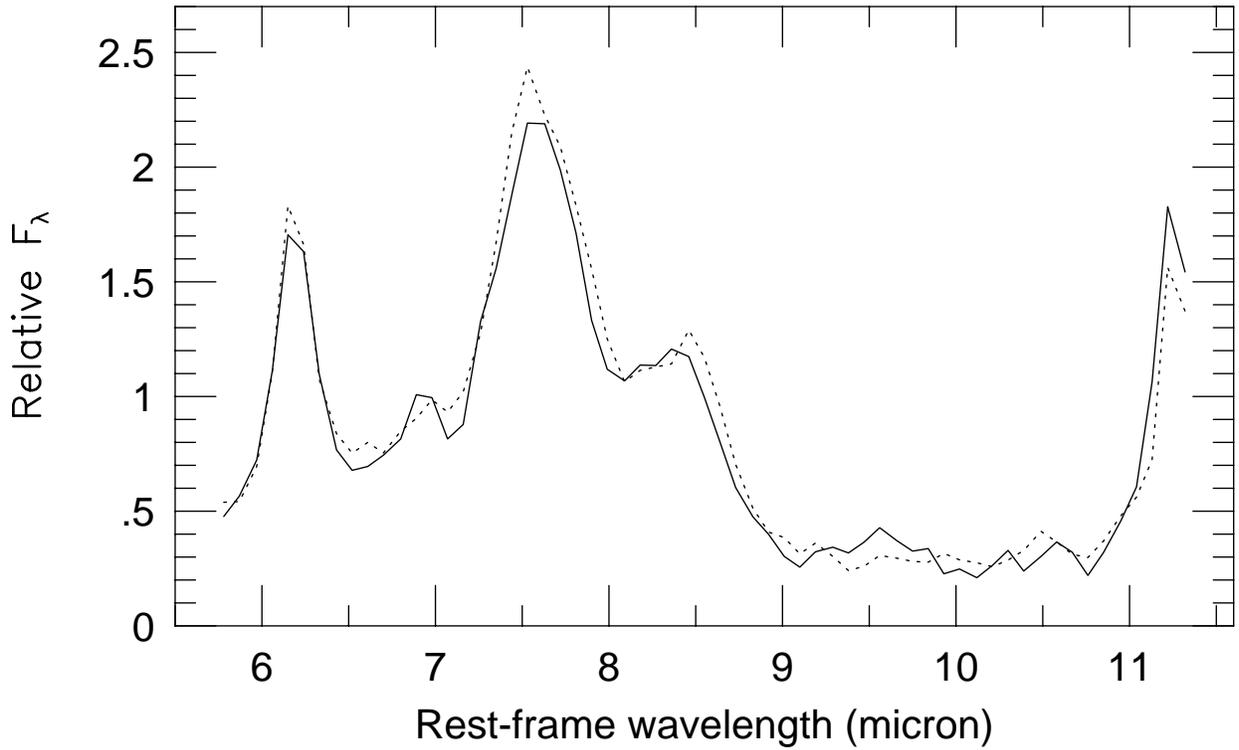,width=16.5cm,bbllx=64pt,bblly=402pt,bburx=523pt,bbury=684pt}}
%  \centerline{\bf\it fig05.ps}
%  \vspace{5.8cm}
%  \end{center}
  \caption{\em  To illustrate the very stable spectral shapes in 
   the PHT-S data,
   Lu \etal\  (2000) co-added eleven FIR-cold galaxies (R(60,100)$<$0.4,
   solid line) 
   and nine FIR-warm galaxies (0.6$<$R(60,100)$>$0.9, dashed line)
   separately, and found very little difference between the two resulting
   class averages.  The only significant difference is a slightly stronger
   11.3$\mu$m feature in the cold galaxies with respect to the 6--9$\mu$m
   features.  The significance of this difference is discussed in the text.
   }
  \label{fig:fig-5}
\end{figure*}

%%%%%%%%%%%%%%%%%%%  FIGURE 6
 \begin{figure*}[b]
%  \begin{center}
%    \leavevmode
  \centerline{\psfig{file=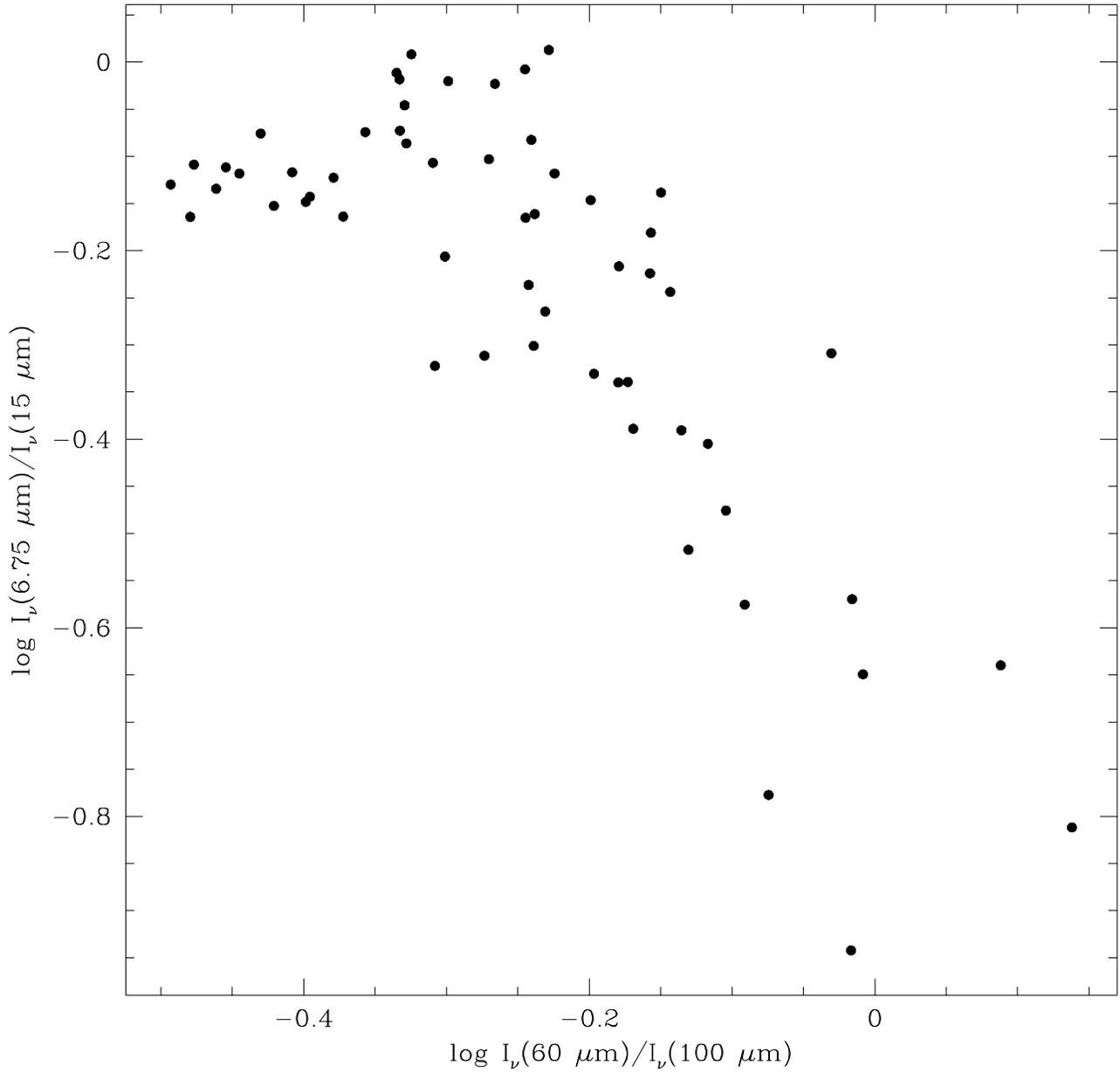,width=18.0cm}}
%  \centerline{\bf\it fig06.ps}
%  \vspace{5.8cm}
%  \end{center}
  \caption{\em The ISO-IRAS color-color diagram for normal star-forming
   galaxies.  The 7\um\ and 15\um\ bands do not show any sign of
   the increased heating signified by the rise of R(60,100) until this latter ratio
   exceeds about 0.6, after which the 15\um band starts detecting the
   rising continuum from warm ``Very Small Grains''.  This plot shows the
   data for the sample used in the ISO Key Project on Normal Galaxies
   (cf. \S 4.1).
   }
  \label{fig:fig-6}

\end{figure*}

%%%%%%%%%%%%%%%%%%%  FIGURE 7
 \begin{figure*}[b]
%  \begin{center}
%    \leavevmode
  \centerline{\psfig{file=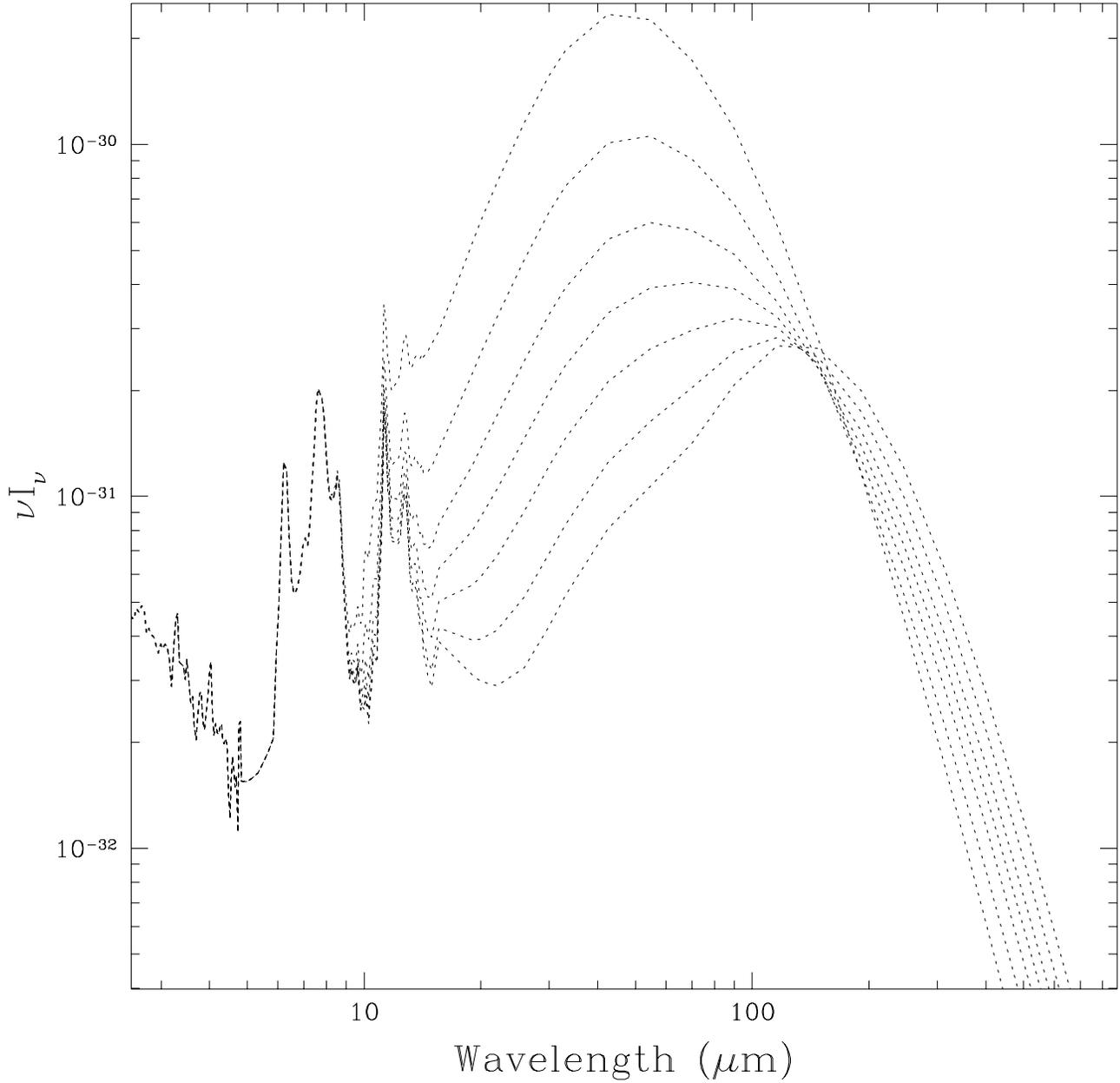,width=18.0cm}}
%  \centerline{\bf\it fig07.ps}
%  \vspace{5.8cm}
%  \end{center}
  \caption{\em  The synthesized model spectra of Dale \etal (2000a), showing 
   the progression of shape from quiescent to active star-forming galaxies.
   The same set is plotted here as summarized in Table~1.
   The spectra are plotted so the Aromatic Features are scaled to the same
   amplitude just for ease of graph-reading, these features being the
   most stable part of the spectra.  Note the march of the peak towards
   shorter wavelengths as activity increases.  These synthesized spectra do
   not account for the general correlation between R(60,100) and the
   infrared-to-visible light ratio.
   }
  \label{fig:fig-7}
\end{figure*}

%%%%%%%%%%%%%%%%% FIGURE 8
\begin{figure}[h]
%  \begin{center}
%    \leavevmode
  \centerline{\psfig{file=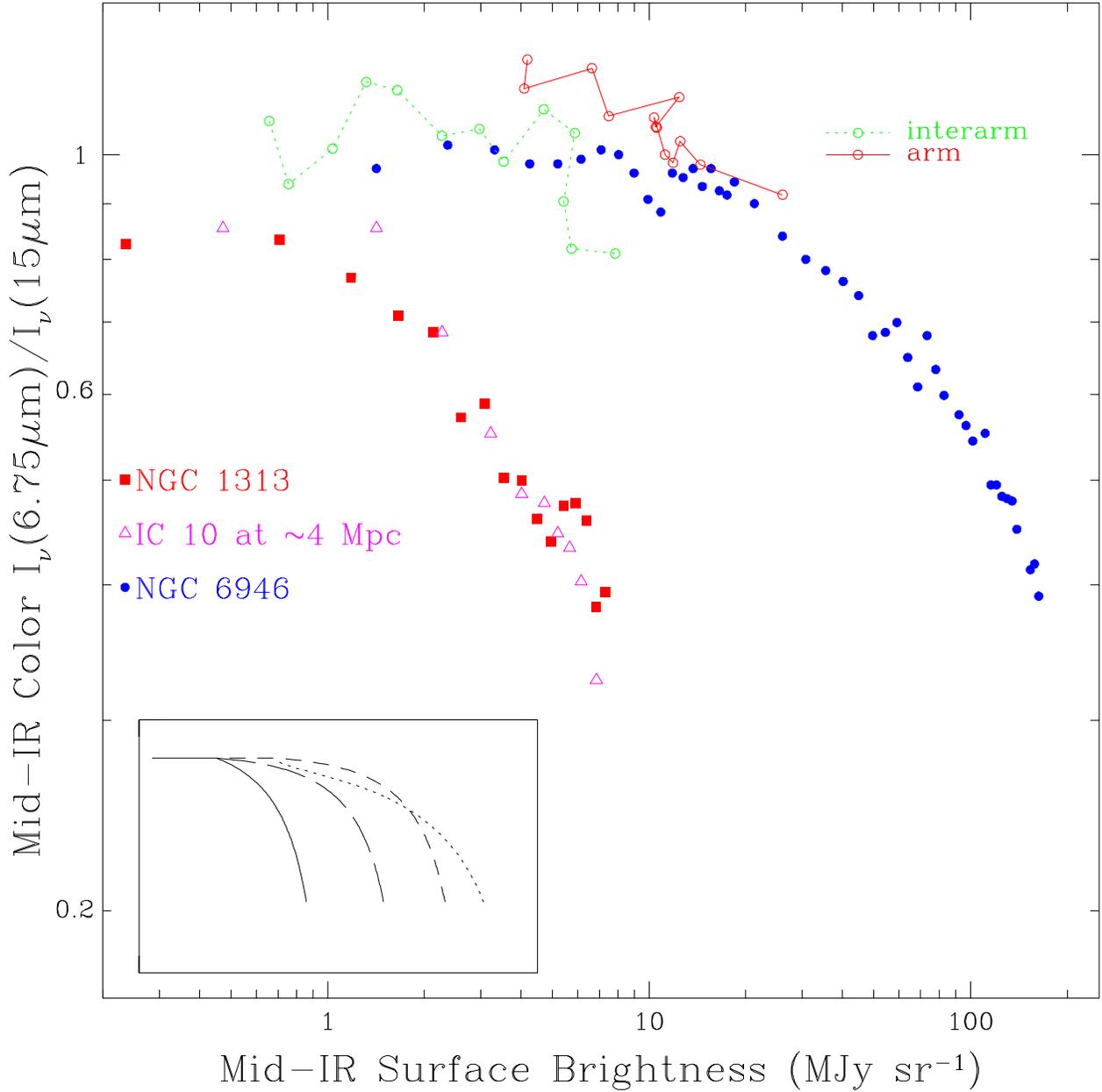,width=18.0cm}}
%  \centerline{\bf\it fig08.ps}
%  \vspace{5.8cm}
%  \end{center}
  \caption{\em The mid-infrared color as a function of surface brightness for
three disk galaxies well resolved by ISOCAM, and smoothed so the resolution
corresponds to $\sim200$~pc in each case (Dale \etal 1999).  
All three data sets show roughly  the same behavior, 
indicative primarily of how color and surface
brightness evolve as heating intensity increases.  The data are consistent
with the expectation that a change in total ISM dust column density will shift
the curves along the surface brightness axis: NGC~6946 does indeed have 
an order of magnitude greater column density than the other 
two galaxies  in the product of $HI+(2/3)H_2$\  and metallicity.}
  \label{fig:fig-8}
\end{figure}

%%%%%%%%%%%%%%%%%%%  FIGURE 9
 \begin{figure*}[b]
%  \begin{center}
%    \leavevmode
  \centerline{\psfig{file=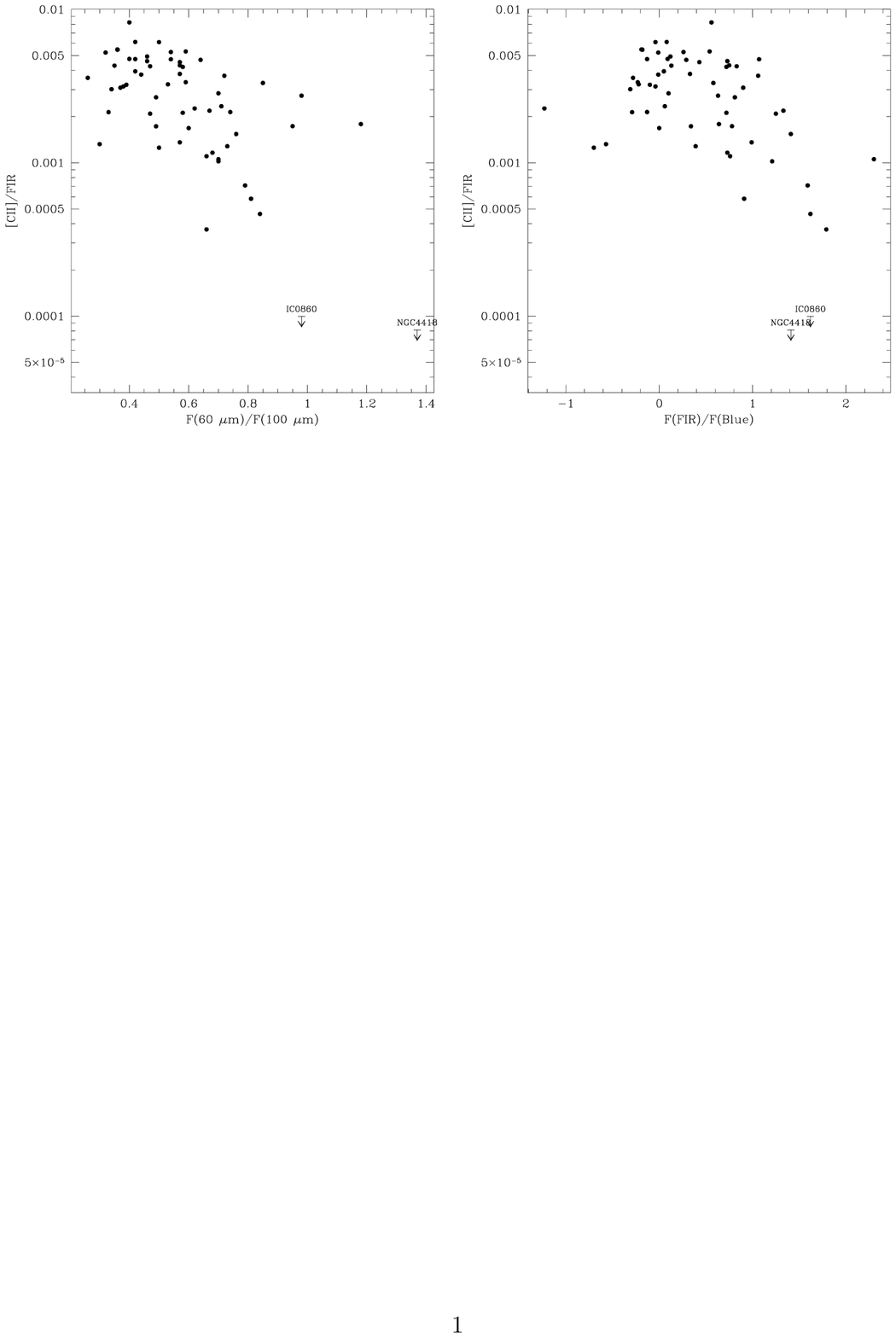,width=16.5cm,bbllx=149pt,bblly=479pt,bburx=560pt,bbury=676pt}}
%  \centerline{\bf\it fig09.ps}
%  \vspace{5.8cm}
%  \end{center}
  \caption{\em The CII deficiency in active star forming galaxies from 
   Malhotra \etal (1997).  The ratio of [CII] to FIR luminosity drops as
   the star formation activity and interstellar heating intensify, as
   measured either by R(60,100) (left-hand side panel), or by the
   infrared-to-visible light ratio (right-hand side panel).  
   See also Malhotra \etal\  1999.
   }
  \label{fig:fig-9}
\end{figure*}

%%%%%%%%%%%%%%%%%%%  FIGURE 10
 \begin{figure*}[b]
%  \begin{center}
%    \leavevmode
  \centerline{\psfig{file=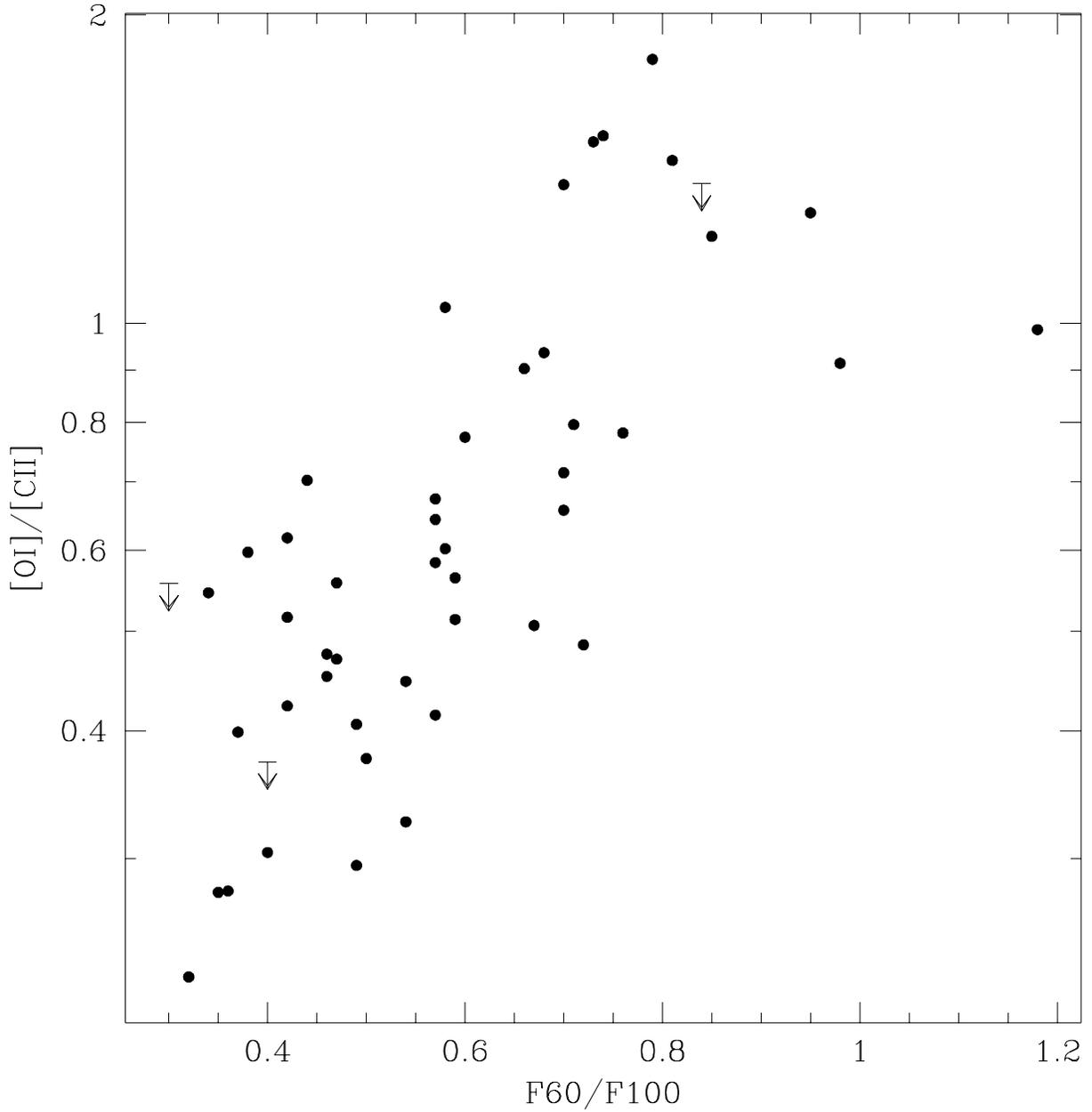,width=18.0cm}}
%  \centerline{\bf\it fig10.ps}
%  \vspace{5.8cm}
%  \end{center}
  \caption{\em The [OI]/[CII] ratio and R(60,100) are positively correlated
   in galaxies
   (Malhotra \etal 1999).  This indicates that the neutral warm gas excitation 
   increases along with the dust heating, so that gas and dust are indeed coupled
   in these galaxies, a result of the photo-electric effect.  The
   quantitative details of the correlation depend on the density and
   heating intensity in the PDRs of these galaxies.  The existence of a
   general correlation for all galaxies indicates that density and heating
   intensity must scale in a similar fashion among all these objects.
   }
  \label{fig:fig-10}
\end{figure*}

\end{document}